%% file: main.tex
\RequirePackage{amsmath}
\documentclass[twocolumn,twocolappendix]{aas}

\input{Lay-out/Preamble}

\begin{document}

\input{Lay-out/Front_matter}

\input{Sections/1_Introduction}
\input{Sections/2_Method}
\input{Sections/3_Results}
\input{Sections/4_Maxwellian}
\input{Sections/5_Comparison}
\input{Sections/6_Conclusions}

\input{Lay-out/Acknowledgements}
\clearpage
\input{Lay-out/Appendix}
\input{Lay-out/References}

\end{document}

%% file: Lay-out/Preamble.tex
\usepackage[utf8]{inputenc}
\usepackage[varg]{txfonts}
\usepackage{gensymb}
\usepackage{nameref}
\usepackage{multirow}
\usepackage{listings}
\usepackage{lipsum}
\usepackage{lmodern}
\usepackage{outlines}
\hypersetup{linkcolor=blue,citecolor=blue,filecolor=blue,urlcolor=blue}

%% file: Lay-out/Front_matter.tex
\title{\Large The Kick Velocity Distribution of Isolated Neutron Stars}

\author[0000-0002-0492-4089]{\normalsize Paul Disberg}
\affiliation{School of Physics and Astronomy, Monash University, Clayton, Victoria 3800, Australia}
\affiliation{The ARC Center of Excellence for Gravitational Wave Discovery---OzGrav, Australia}
\email[show]{\href{mailto:paul.disberg@monash.edu}{paul.disberg@gmail.com}}

\author[0000-0002-6134-8946]{\normalsize Ilya Mandel}
\affiliation{School of Physics and Astronomy, Monash University, Clayton, Victoria 3800, Australia}
\affiliation{The ARC Center of Excellence for Gravitational Wave Discovery---OzGrav, Australia}
\email{ilya.mandel@monash.edu}

\received{May 28, 2025}
\revised{July 3, 2025}
\accepted{July 21, 2025}
\submitjournal{ApJL}

\begin{abstract}
\noindent Neutron stars (NSs) are thought to receive natal kicks at their formation in supernovae. In order to investigate the magnitude of these kicks, we analyze the proper motions and distance estimates---either through parallax or dispersion measures---of young isolated pulsars and infer their three-dimensional velocities relative to their local standard of rest. We find that the velocities based on parallax distances of pulsars younger than $10$ Myr follow a log-normal distribution with $\mu=5.60\pm0.12$ and $\sigma=0.68\pm0.10$, peaking at ${\sim}150$--$200$ km s$^{-1}$, which we adopt as our fiducial kick distribution. Using a previously established method that infers kick magnitudes through the eccentricity of Galactic trajectories, we also estimate the kick velocities of older pulsars, which we find to be consistent with our fiducial kick distribution. A log-normal fit to all pulsars with ages below $40$ Myr yields a more constraining (but possibly more prone to systematic errors) fit with $\mu=5.67\pm0.10$ and $\sigma=0.59\pm0.08$, respectively. Moreover, we (1) resolve the tension between our results and the Maxwellian distribution found by Hobbs et al.\ (2005), which has a ${\sim}50\%$ higher median velocity, by showing that their analysis is missing a Jacobian needed to correct for its logarithmic histogram bin sizes, and (2) argue that the bimodality found by others is not statistically significant and that previous results are consistent with our inferred kick distribution, effectively reconciling the literature on observed NS kicks. 
\end{abstract}

%% file: Sections/1_Introduction.tex
\section{Introduction}
\noindent Neutron stars (NSs) are born in supernovae which are likely anisotropic \citep[e.g.,][]{Janka_1994,Burrows_1995,Herant_1995}, causing them to receive natal kick velocities \citep[e.g.,][]{VandenHeuvel_1997} with magnitudes of ${\sim}100$--$1000$ km s$^{-1}$ \citep{Burrows_2024}. NSs observed as isolated pulsars are indeed known to have higher velocities than their progenitors \citep[e.g.,][]{Gunn_1970,Lyne_1994}, but the distribution of the kick velocities that these objects receive remains a topic of discussion. Several studies have tried to constrain the natal kicks of NSs based on the observed properties of isolated pulsars \citep[e.g.,][]{Arzoumanian_2002,Hobbs_2005,Faucher_2006,Verbunt_2017,Igoshev_2020,Disberg_2025}, whereas other studies have investigated the natal kicks of NSs in binaries \citep[e.g.,][]{Fortin_2022,Zhao_2023}. These empirical natal kick distributions are often used in population synthesis models of binary evolution.

In particular, \citet{Hobbs_2005} used the proper motion of young isolated pulsars to infer the three-dimensional velocities that they are a projection of. \citet{Hobbs_2005} argued that these are well-described by a Maxwellian distribution with $\sigma = 265$ km s$^{-1}$ (peaking at a three-dimensional speed of ${\sim}400$ km s$^{-1}$). Since these pulsars are young (i.e., younger than $3$ Myr) their velocities are indicative of the kick magnitude. 

However, \citet{Verbunt_Cator_2017} argue that there are systematic errors in the distance estimates used by \citet{Hobbs_2005} that are based on dispersion measures (DM) and electron density models \citep[e.g.,][]{Cordes_1998,Cordes_2002,Yao_2017}, which affect their velocity estimates. Indeed, they show that the transverse velocities of the pulsars analyzed by \citet{Brisken_2002} are difficult to explain with the kick distribution of \citet{Hobbs_2005}.

For this reason, \citet{Verbunt_2017} inferred the three-dimensional velocities of pulsars that have distance estimates determined through parallax (as opposed to DM), and found that these velocities are best described by a weighted sum of two Maxwellians. However, they note that this does not necessarily mean that the kick distribution is truly bimodal, since it could also be explained by a distribution that is wider than a single Maxwellian. \citet{Igoshev_2020}, in turn, redid the analysis of \citet{Verbunt_2017} after significantly expanding their pulsar sample with the parallax estimates of \citet{Deller_2019}, and also found a kick distribution best described by two Maxwellians \citep[see also][]{Igoshev_2021}, similarly to the results of \citet{Verbunt_2017}. The kick distributions of \citet{Verbunt_2017} and \citet{Igoshev_2020} favor significantly lower velocities than the Maxwellian found by \citet{Hobbs_2005}.

Recently, \citet{Disberg_2024a} investigated the transverse and three-dimensional velocities of pulsars, and found that older pulsars have lower velocities since their kicks cause their Galactic trajectories to be more eccentric \citep[cf.][]{Hansen_1997}. Subsequently, \citet{Disberg_2024b} used the fact that larger kicks lead to more eccentric Galactic trajectories to determine the systemic kicks of BNSs. That is, they estimated the eccentricities of Galactic BNSs and used a simulation of kicked objects moving in the Galactic potential to determine the relationship between kick magnitude and eccentricity, from which the (systemic) kicks of the BNSs could be inferred \citep[cf.\ the method of][]{Atri_2019}.

This method was expanded by \citet{Disberg_2025} and applied to the pulsar samples---including distance estimates---of \citet{Igoshev_2020}, \citet{Verbunt_2017}, and \citet{Hobbs_2005}. While they could recreate the findings of \citet{Igoshev_2020} and \citet{Verbunt_2017} for young pulsars (including bimodality), they found that their results for the (young) sample of \citet{Hobbs_2005} contained significantly lower kicks than predicted by the Maxwellian with $\sigma=265$ km s$^{-1}$, instead following a distribution more similar to the other samples. Moreover, taking into account both young and older pulsars, \citet{Disberg_2025} found that the NS kicks were best described by a log-normal distribution, and conjectured that the bimodality found for young pulsars might be caused by a limited sample size. Since \citet{Disberg_2025} used the DM distance estimates of \citet{Hobbs_2005}, they argued it is difficult to explain the tension between the Maxwellian kick distribution of \citet{Hobbs_2005} and the findings of \citet{Verbunt_2017}, \citet{Igoshev_2020}, and \citet{Disberg_2025} with systematic errors in the distance estimates.

In this work, we use the largest available sample of both young and older pulsars with either DM or parallax distances. We (1) analyze the current velocities of young pulsars, (2) use the method proposed by \citet{Disberg_2024b}---and employed by \citet{Disberg_2025}---to kinematically constrain the kicks of old pulsars, (3) investigate the tension in the literature, and (4) fit a consistent model to the distribution of pulsar velocities. In Section \ref{sec2}, we describe the method used to analyze the kicks of the pulsars in our sample, through distance estimates based on parallax or DM. We use the parallax distances in particular to determine a fiducial kick distribution, which we show in Section \ref{sec3}. In Section \ref{sec4}, we investigate the Maxwellian distribution found by \citet{Hobbs_2005} and argue that it is likely based on an erroneous histogram interpretation due to a missing Jacobian. We apply our method to the pulsar samples of \citet{Igoshev_2020}, \citet{Verbunt_2017}, and \citet{Hobbs_2005}, and in Section \ref{sec5} we show that all observations are consistent with our fiducial kick distribution. Finally, in Section \ref{sec6} we summarize our conclusions.

%% file: Sections/2_Method.tex
\section{Method}
\label{sec2}
\renewcommand{\thefootnote}{1}
\noindent We are interested in analyzing the current three-dimensional velocities of young pulsars and applying the method of \citet{Disberg_2024b,Disberg_2025} to old pulsars. In particular, we consider the pulsars from the Australia Telescope National Facility (ATNF) Pulsar Catalogue\footnote{\href{https://www.atnf.csiro.au/research/pulsar/psrcat/}{https://www.atnf.csiro.au/research/pulsar/psrcat/}.} v$2.6.0$ \citep{Manchester_2005}, and only select pulsars with (1) a distance estimate through DM which does not exceed $10$ kpc, (2) a proper motion estimate, (3) a characteristic spin-down age estimate (i.e., $\tau_{\text{c}}=1/2\cdot P/\dot{P}$), (4) $\dot{P}>5\cdot10^{-18}$, to exclude recycled pulsars, (5) no evidence for the pulsar being part of a binary system, (6) no association with a globular cluster, and (7) no classification as either an \textquotedblleft Anomalous X-ray Pulsar or Soft Gamma-ray Repeater with detected pulsations\textquotedblright\ (AXP) or an \textquotedblleft Isolated Neutron Star with pulsed thermal X-ray emission but no detectable radio emission\textquotedblright\ (XINS). This selection results in a sample of $197$ pulsars with DM distance estimates, of which $76$ have parallax distance estimates. These parallax estimates update and expand the pulsar sample used by \citet{Igoshev_2020}---which contains the results of \citet{Chatterjee_2001}, \citet{Brisken_2002}, \citet{Brisken_2003}, \citet{Chatterjee_2004}, \citet{Chatterjee_2009}, \citet{Deller_2009}, \citet{Kirsten_2015}, and \citet{Deller_2019}---with the measurements of \citet{Bruzewski_2023}, \citet{Lin_2023}, and \citet{Keith_2024}. Moreover, in Appendix \ref{appA} we show the timing parameters (i.e., $P$ and $\dot{P}$) of the pulsars in our sample and the distribution of their characteristic ages.

In order to describe the probability distribution for a parallax distance estimate (i.e., $D_{\pi}$) corresponding to a certain observed parallax (i.e., $\omega_{\text{obs}}$) and its uncertainty (i.e., $\sigma_{\omega}$), we use the formalism proposed by \citet{Verbunt_Cator_2017} and employed by \citet{Verbunt_2017} and \citet{Igoshev_2020}. That is, the posterior distribution of $D_{\pi}$ given an observed parallax $\omega_{\text{obs}}$ is given by \citep[see also][]{Lorimer_2006,Verbiest_2012,Igoshev_2016}:
\begin{multline}
    \label{eq1}
    p(D_{\pi}\text{\hspace{.4mm}}|\text{\hspace{.4mm}}\omega_{\text{obs}})\propto\exp\left(-\frac{\left(\left[D_{\pi}/\text{kpc}\right]^{-1}-\left[\omega_{\text{obs}}/\text{mas}\right]\right)^2}{2\left[\sigma_\omega/\text{mas}\right]^2}\right)\\ \times D_{\pi}^2R^{1.9}\exp\left(-\frac{|z|}{0.33\ \text{kpc}}-\frac{R}{1.70\ \text{kpc}}\right),
\end{multline}
where $z$ and $R$ are the galactocentric cylindrical coordinates of the pulsar, where $z=D_{\pi}\sin b$ and $R^2=R_{\sun}^2+\left(D_{\pi}\cos b\right)^2-2R_{\sun}D_{\pi}\cos b\cos l$ for Galactic coordinates $l,\,b$ and galactocentric radius of the Solar System $R_{\sun}=8.122$ kpc \citep{Gravity_2018}. We determine the parallax distances (for each parallax pulsar) by sampling $10^3$ values from Equation \ref{eq1}, whereas the DM distances (for all pulsars) are estimated by sampling $10^3$ values from a normal distribution centered around the distance determined through the electron density model of \citet{Yao_2017}, with a standard deviation of $20\%$. Although detailed studies such as those of \citet{Deller_2009} show that DM distances can sometimes have even greater systematic errors, we use an indicative error of $20\%$ \citep[e.g.][]{Ding_2024}. We note that we primarily use DM distance estimates to show that these are not sufficient to explain the differences between the results of \citet{Verbunt_2017} and those of \citet{Hobbs_2005}.

Having obtained $10^3$ DM distance samples for all $197$ pulsars and $10^3$ parallax distances for the $76$ pulsars with parallax estimates, we also sample $10^3$ proper motion samples from Gaussians based on the observed values and their uncertainties. Then, we convert the coordinates to the galactocentric frame (essentially adding the velocity of the Solar System), and project the resulting velocity vectors on the line of sight and the direction perpendicular to the line of sight. The latter we define as the galactocentric transverse velocity $v_{t}$ \citep[see also][]{Gaspari_2024a}. If the galactocentric velocity vector $\vec{v}$ is projected on the line of sight and the plane of the sky, it can be described as:
\begin{equation}
    \label{eq2}
    \vec{v}=\begin{pmatrix}v_{\alpha}\\v_{\delta}\\v_r\end{pmatrix}=v\begin{pmatrix}\sin\phi\sin\theta\\\cos\phi\sin\theta\\\cos\theta\end{pmatrix},
\end{equation}
where the transverse velocity equals $v_t=\sqrt{v_{\alpha}^2+v_{\delta}^2}=v\sin\theta$ and the radial velocity equals $v_r=v\cos\theta$ (note that \textquotedblleft radial\textquotedblright\ here refers to the line of sight, rather than the Galactic radial coordinate). If this vector is oriented isotropically in the galactocentric frame (GC isotropy), then $\phi$ is distributed uniformly and $p(\theta)=1/2\cdot\sin\theta$ for $0\leq\theta\leq\pi$. This means that for GC isotropy, the radial component of the velocity vector is given by
\begin{equation}
    \label{eq3}
    v_r=v_t\hspace{.3mm}\cot\theta\quad\left(\text{GC isotropy}\right),
\end{equation}
where $\theta=\arccos u$ (with $u$ being uniformly sampled between $-1$ and $1$), since $p(\theta)\text{d}\theta=p(u)\text{d}u$ and $p(u)=p(-u)=1/2$. However, if the velocity vector is oriented isotropically in the local standard of rest (LSR) of the pulsar, the galactocentric radial velocity is \citep{Gaspari_2024a}:
\begin{equation}
    \label{eq4}
    v_r=\left|\left|\vec{v}_t-\vec{v}_{\text{LSR},\,t}\right|\right|\hspace{.2mm}\cot\theta+v_{\text{LSR},\,r}\quad\left(\text{LSR isotropy}\right),
\end{equation}
where $\vec{v}_{\text{LSR},\,t}$ and $v_{\text{LSR},\,r}$ describe the transverse and radial velocities of the LSR of the pulsar relative to the Solar System.

\input{Lay-out/Table1}

We combine the distance estimates with the proper motions to create samples of the transverse velocity for each pulsar, and then use Equation \ref{eq4} to create a corresponding sample for the radial velocity distribution, since LSR isotropy is a more accurate assumption than GC isotropy \citep{Disberg_2024b}. The transverse and radial velocity samples form the three-dimensional velocity vectors in the galactocentric frame, which we transformed to velocities relative to the pulsars' LSR by subtracting $\vec{v}_{\text{LSR}}$. We approximate $v_{\text{LSR}}$ by using the circular velocity at the location of the pulsar \citep[as given by the Milky Way potential of][]{McMillan_2017}. As a result we obtain the present-day velocities of the pulsars relative to their LSR---which we define as \textquotedblleft$v_{\text{pd}}$\textquotedblright. For pulsars with characteristic spin-down ages $\tau_{\text{c}}$ below $10$ Myr, we approximate their kicks by estimating $v_{\text{pd}}$, since for pulsars with these ages their current velocities are still representative of their kicks \citep{Disberg_2024a}. Our sample contains 142 pulsars with $\tau_{\text{c}} < 10$ Myr, of which $48$ have parallax estimates. Moreover, although there are cases in which $\tau_{\text{c}}$ does not appear to be accurate (e.g., \citealt{Lyne_1996} find a factor $2$--$3$ difference between $\tau_{\text{c}}$ and their age estimates), it tends to agree with kinematic age estimates \citep[e.g.,][]{Noutsos_2013,Igoshev_2019,Disberg_2025} and in particular \citet{Maoz_2024} argue that it is a reasonable age estimate.

\renewcommand{\thefootnote}{2}
For pulsars older than $10$ Myr, their present-day velocities have increasingly been affected by their migration through the Galactic potential \citep{Disberg_2024a}, which is why we employ the method of \citet{Disberg_2025} to estimate the kicks of these older pulsars. That is, we sample $10^2$ velocity vectors for each of these pulsars, similarly to the young ones, trace their trajectories back through the Galaxy \citep[with \lstinline{GALPY}\footnote{\href{http://github.com/jobovy/galpy}{http://github.com/jobovy/galpy}.} v$1.10.0$,][]{Bovy_2015}, and estimate the eccentricity of their Galactic orbits, since eccentricity is a good indicator of kick magnitude even for older pulsars \citep{Disberg_2024b}. Then, we use the simulated relationship between eccentricity and kick magnitude found by \citet[][see their Figure 4]{Disberg_2025}. This relationship describes the Galactic eccentricities of kicked objects within the solar neighborhood (defined as distances below $2$ kpc), integrated over a given time domain. For each of the $100$ eccentricity values per pulsar we sample $10$ kick velocities from the relationship between kick velocity and Galactic eccentricity, while assuming a flat prior on the kicks, and obtain a distribution of kinematically constrained kick velocities---which we define as \textquotedblleft$v_{\text{kc}}$\textquotedblright---where we correct for detection bias by considering the fraction of objects that cross the solar neighborhood as a function of kick magnitude.

The benefit of this method is that it can be used to obtain information about the kicks of old objects, whereas previous methods used to determine NS kicks only analyzed the current velocities of young objects. However, the downside of this method is that this model makes several assumptions that bring in more systematic uncertainties than the method used for young pulsars. For example, we integrate the eccentricity versus kick relationship over a time domain and compare it to pulsars with ages within this domain, but this relationship evolves over time and the distribution of pulsar ages is generally not uniform within a given time domain (see Appendix \ref{appA}). The boundaries of the time domains are sensitive to the accuracy of $\tau_{\text{c}}$ as an approximation for the pulsar age. We assume a static Milky Way model because Galactic evolution since the formation of the pulsars is not significant \citep[e.g.][]{Amend_2024}. Moreover, our model of the detection probability of a kicked neutron star, through the probability that it reaches (our definition of) the solar neighborhood, is an approximation that brings in systematic uncertainty as well. Because of these systematic model uncertainties, we base our fiducial kick distribution only on the young pulsars (i.e., $\tau_{\text{c}}<10$ Myr).

%% file: Lay-out/Table1.tex
\begin{table*}
\caption{Parameters of the log-normal fits (Equations \ref{eq5} and \ref{eq6}) to the velocity distributions of the pulsars in our sample.\label{tab1}}
\hspace{-29mm}\begin{tabular}{l|ccccc|ccccc}
\hline\hline\\[-13pt]
age range & \multicolumn{5}{c|}{parallax distances} & \multicolumn{5}{c}{DM distances}\\
 & \multirow{2}{*}{$\mu$} & \multirow{2}{*}{$\sigma$} & \multirow{2}{*}{$N$} & median & mode & \multirow{2}{*}{$\mu$} &\multirow{2}{*}{$\sigma$} & \multirow{2}{*}{$N$} & median & mode \\[-2pt]
 & & & & $\left[\text{km}\,\text{s}^{-1}\right]$ & $\left[\text{km}\,\text{s}^{-1}\right]$ & & & & $\left[\text{km}\,\text{s}^{-1}\right]$ & $\left[\text{km}\,\text{s}^{-1}\right]$ \\
[2pt]\hline\\[-13pt]
$\tau_{\text{c}}\,{\leq}\,3\,\text{Myr}$ &5.56(17)&0.79(12)&29&248&139&5.39(09)&0.93(07)&94&210&92\\
$\tau_{\text{c}}\,{\leq}\,10\,\text{Myr}$ &\textbf{5.60(12)}&\textbf{0.68(10)}&\textbf{48}&\textbf{266}&\textbf{170}&5.29(09)&0.85(07)&142&190&96\\
$10\,\text{Myr}\,{<}\,\tau_{\text{c}}\,{\leq}\,40\,\text{Myr}$ &5.57(24)&0.35(19)&17&262&232&5.70(39)&0.55(25)&33&295&221\\
$\tau_{\text{c}}\,{\leq}\,40\,\text{Myr}$ &5.67(10)&0.59(08)&65&287&205&5.39(08)&0.79(06)&175&212&117\\
[2pt]\hline
\end{tabular}
\tablecomments{\footnotesize The values in parentheses are half the widths of the $68\%$ credible intervals, $N$ is the sample size, and the bold values show our fiducial kick distribution. The median values are determined for the distribution between $0$ and $1000$ km s$^{-1}$.}
\end{table*}

%% file: Sections/3_Results.tex
\begin{figure*}
    \centering
    \includegraphics[width=18cm]{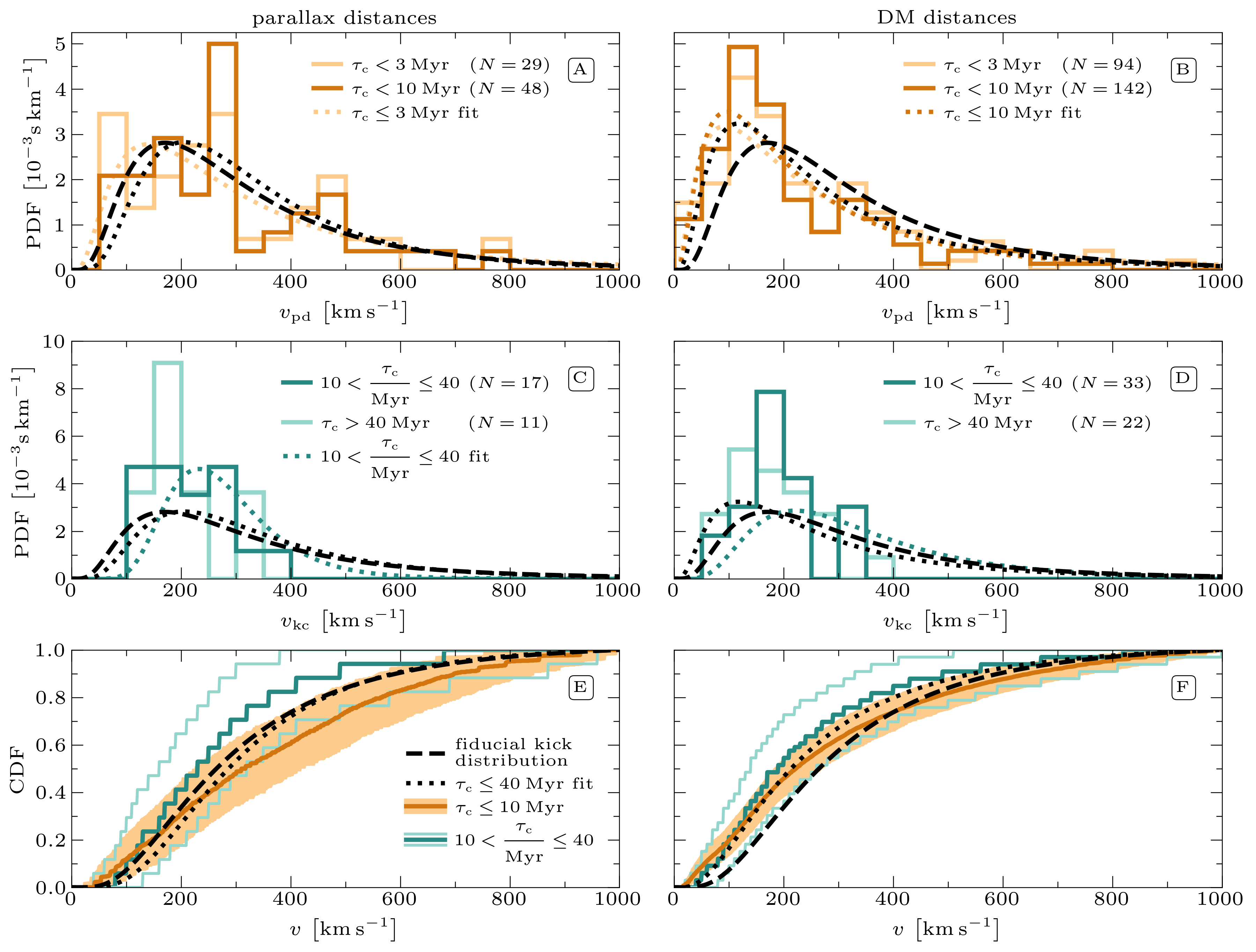}
    \caption{Kick estimates for our pulsar sample, for parallax distances (left column) and DM distances (right column). The top row panels show the distributions of the most likely present-day velocities ($v_{\text{pd}}$) of all pulsars younger than $3$ Myr (light brown) and younger than $10$ Myr (brown), in histograms with bins of $50$ km s$^{-1}$. The dashed black line in all panels is the fiducial kick distribution resulting from a log-normal fit (Equations \ref{eq5}, $\mu=5.60$ and $\sigma=0.68$) to the $v_{\text{pd}}$ distribution based on parallax distances for $\tau_{\text{c}}\leq10$ Myr. The middle row panels shows a histogram of the most likely values of the kinematically constrained kicks $v_{\text{kc}}$ for pulsars with ages between $10$ and $40$ Myr (teal) and pulsars older than $40$ Myr (light teal). In all panels, the dotted lines show the log-normal fits listed in Table \ref{tab1}, where the black dotted line corresponds to the fit for $\tau_{\text{c}}\leq40$ Myr. The top two rows also display the number of pulsars in each subsample ($N$). The bottom row panels show CDFs of the (median) $v_{\text{pd}}$ empirical distribution for $\tau_{\text{c}}\leq10$ Myr (brown line) and the (median) $v_{\text{kc}}$ distribution for $10<\tau_{\text{c}}/\text{Myr}\leq40$ (teal line). We estimated the $95\%$ confidence intervals of these distributions by bootstrapping them (light brown area and light teal lines, respectively). The CDFs in the bottom panels are normalized between $0$ and $1000$ km s$^{-1}$.}
    \label{fig1}
\end{figure*}

\section{Results}
\label{sec3}
\noindent We divide our pulsar sample into four age ranges: $\tau_{\text{c}}\leq3$ Myr \citep[similarly to][]{Hobbs_2005,Igoshev_2020}, $\tau_{\text{c}}\leq10$ Myr \citep[similarly to][]{Verbunt_2017}, $10$ Myr $ < \tau_{\text{c}}\leq40$ Myr, and $\tau_{\text{c}}>40$ Myr, and sample $10^3$ present-day velocity values for the pulsars younger than $10$ Myr and $10^3$ kinematically constrained kick values for the ones older than $10$ Myr. We fit a log-normal distribution \citep[similarly to][]{Disberg_2025} to the kick distributions for each age range:
\begin{equation}
    \label{eq5}
    p_{\text{ln}}(v\text{\hspace{.4mm}}|\text{\hspace{.4mm}}\mu,\sigma)=\frac{1}{v\sigma\sqrt{2\pi}}\exp\left(-\frac{\left(\ln v-\mu\right)^2}{2\sigma^2}\right),
\end{equation} 
which we renormalize for the velocity range $0 \leq v \leq 1000$ km s$^{-1}$.
The likelihood of obtaining the set of observations given parameters $\mu$ and $\sigma$ is \citep[see][]{Mandel_2019}:
\begin{equation}
    \label{eq6}
    \mathcal{L}\left(\mu,\sigma\text{\hspace{.4mm}}|\text{\hspace{.4mm}}\vec{v}\right)=\prod_{n=1}^{N}\dfrac{1}{10^3}\sum_{i=1}^{10^3}\dfrac{p_{\text{ln}}(\vec{v}_{n,i}\text{\hspace{.4mm}}|\text{\hspace{.4mm}}\mu,\sigma)}{\int p_{\text{det}}(v)\,p_{\text{ln}}(v\text{\hspace{.4mm}}|\text{\hspace{.4mm}}\mu,\sigma)\,\text{d}v},
\end{equation}
where $\vec{v}$ is the vector containing the resulting velocity values such that $\vec{v}_{n,i}$ is the $i^{\text{th}}$ sample for the $n^{\text{th}}$ pulsar (for a data set containing $N$ pulsars), and $p_{\text{det}}(v)$ is the probability of detecting a pulsar that received a kick equal to $v$, which is assumed to be uniform for young pulsars and for old pulsars is determined by the simulation of \citet[][see their Figure 4]{Disberg_2025}. We fit the log-normal distribution by finding the values of $\mu$ and $\sigma$ for which the likelihood $\mathcal{L}$ is maximal.

In Table \ref{tab1} we list the log-normal parameters (i.e., $\mu$ and $\sigma$) fitted to pulsar velocities in the various age ranges. The younger the pulsars, the more representative their current velocities are for their natal kicks, but in order to limit the effects of low number statistics we consider the fit to pulsars with $\tau_{\text{c}}\leq10$ Myr (based on parallax distances), which results in $\mu=5.60\pm0.12$ and $\sigma=0.68\pm0.10$, to be our fiducial kick distribution. For the oldest pulsars with $\tau_{\text{c}}>40$ Myr, we find the data to be insufficient for constraining $\mu$ and $\sigma$. However, through Equation \ref{eq6} we are able to fit a log-normal distribution to the pulsars with ages between $10$ and $40$ Myr, and by multiplying the likelihood for this sample with the likelihood for pulsars with $\tau_{\text{c}}\leq10$ Myr we constrain the parameters even further and obtain a fit for the pulsars with $\tau_{\text{c}}\leq40$ Myr, which results in $\mu=5.67\pm0.10$ and $\sigma=0.59\pm0.08$ (for parallax distance estimates).

The statistical uncertainties on $\mu$ and $\sigma$ quoted in Table \ref{tab1} are half the width of the $68\%$ credible intervals on these quantities obtained from the posterior on $\mu$ and $\sigma$ assuming the likelihood function of Equation \ref{eq6} and flat priors. We note that for parallax distances our parameter estimates are consistent within their error bars, though the estimated $\sigma$ for the $10<\tau_{\text{c}}/\text{Myr}\leq40$ sample is a marginal outlier. Table \ref{tab1} also lists fits based on DM distance estimates, but we are more confident in the distance estimates based on parallax. The parallax estimates are marginally consistent with the parameter estimates for DM distances, which favor smaller $\mu$ and wider $\sigma$.

In Figure \ref{fig1} we show log-normal fits for the DM distances and the parallax distances, together with histograms of the most likely velocity values for each pulsar (after reweighting by our fiducial kick distribution as the prior). The top row shows the velocity estimates of young pulsars (for $\tau_{\text{c}}\leq3$ Myr and $\tau_{\text{c}}\leq10$ Myr), relative to the LSR of the pulsar, assuming LSR isotropy (Equation \ref{eq4}). These distributions appear to be well-described by our fiducial kick distribution as well as our combined fit for all pulsars with $\tau_{\text{c}}\leq40$ Myr, both of which peak at ${\sim}150$--$200$ km s$^{-1}$. Both of these are in general agreement with the distribution found by \citet{Disberg_2025}, although the current analysis is able to better constrain the high-velocity tail of the distribution. Panel B shows the corresponding velocity distributions based on the DM distances, which contain slightly lower velocities than the parallax-based velocities. These distributions still follow log-normal distributions, which peak at ${\sim}100$ km/s.

The second row of Figure \ref{fig1} shows the kinematically constrained kick velocities for the older pulsars. The resulting distributions generally agree with the velocity distribution for the young pulsars: they are centered around ${\sim}200$ km s$^{-1}$ for the parallax distances, and contain slightly lower kicks for the DM distances. The log-normal fit to the pulsars with ages between $10$ Myr and $40$ Myr is in general agreement with our fiducial kick distribution: it peaks at ${\sim}200$ km s$^{-1}$.

The last row of Figure \ref{fig1} shows cumulative distribution functions (CDFs) of the velocity estimates for pulsars with $\tau_{\text{c}}\leq10$ Myr and the kinematically constrained kick estimates for $10<\tau_{\text{c}}/\text{Myr}\leq40$. The $95\%$ observational confidence intervals are obtained by bootstrapping: we randomly select (with replacement) $N$ pulsars from the data set and for each pulsar select one velocity sample to create a single observational CDF, then repeat this $10^3$ times and determine the $95\%$ intervals of the resulting set of empirical CDFs. For the parallax distance estimates (panel E), both the fiducial kick distribution and the fit for all pulsars with $\tau_{\text{c}}\leq40$ Myr lie within the confidence intervals for both young and older pulsars, despite the fact that the older pulsar sample is likely biased against high kicks. A log-normal distribution is a sufficient model for the kicks of the pulsars in our data set, and we find no evidence for a statistically significant bimodality \citep[cf.][]{Verbunt_2017,Igoshev_2020} or multimodality \citep[cf.][]{Valli_2025} in the kick distribution. It is unsurprising that the log-normal fit does not trace the median velocity distribution perfectly because the velocity distributions for the individual pulsars are asymmetrically skewed to higher values, mainly due to the projection factor of $\cot\theta$ in Equations \ref{eq3} and \ref{eq4}, as we show in Appendix \ref{appB}. 

In our fiducial kick distribution only ${\sim}1\%$ of pulsars obtain low kicks \citep[below $50$ km s$^{-1}$, following][]{Willcox_2021}, which are relevant for retaining NSs in globular clusters or wide binaries, although in the $95\%$ confidence region this fraction may be as large as ${\sim}10\%$. It seems likely that there is a subpopulation of NSs that are formed following binary interactions, receive low natal kicks, therefore remain in binaries such as low-eccentricity Be X-ray binaries \citep[e.g.,][]{Valli_2025}, wide low-eccentricity detached NS binaries observed with Gaia (though post-natal kick mechanisms could be at play, see \citealt{Hirai_2024}) and in clusters \citep[e.g.,][]{Smits:2006}, and are thus explicitly excluded from our observational isolated pulsar data set. These might, for example, originate from electron-capture supernovae that happen predominantly or exclusively for progenitors stripped by mass transfer and thus avoiding second dredge-up \citep{Willcox_2021} or from accretion induced collapse. Another possibility to consider is that NSs that receive low kicks are somehow different from typical pulsars and avoid pulsar surveys unless they are recycled, but there is no evidence for a correlation between kick velocity and pulsar detectability in the existing data \citep{Willcox_2021}.

\begin{figure*}
    \centering
    \includegraphics[width=18cm]{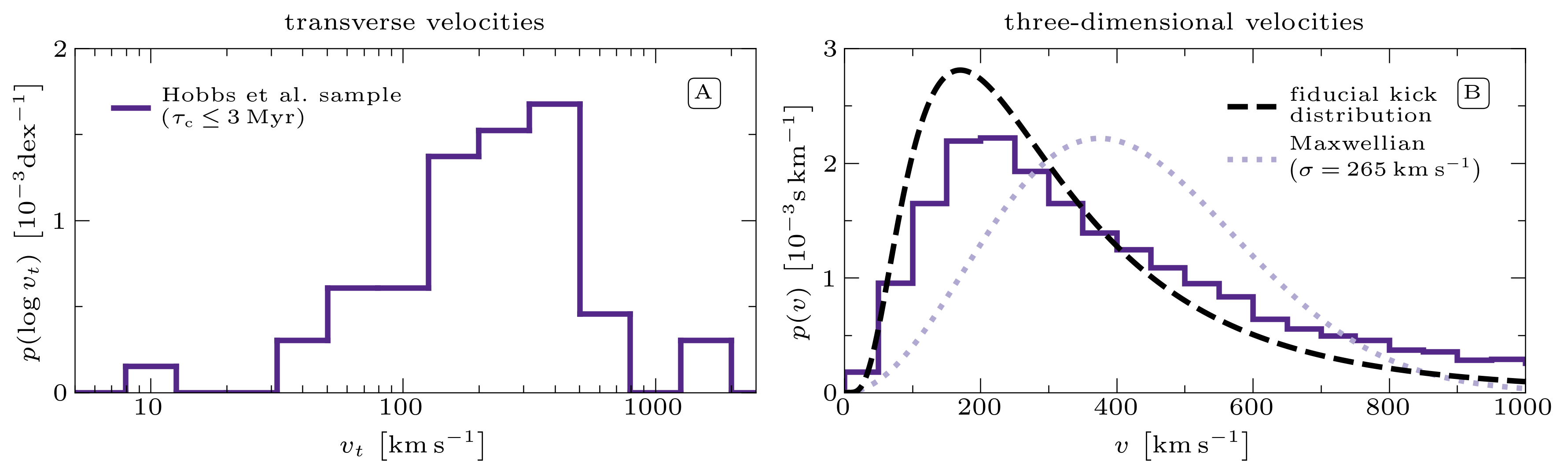}
    \caption{Analysis of the young (i.e., $\tau_{\text{c}}\leq3$ Myr) pulsar sample of \citet{Hobbs_2005}. Panel A contains a histogram of the logarithms of (observer frame) transverse velocities ($\log v_{t}$) in uniform logarithmic bins. This histogram is identical to the one of \citet[][see their Figure 4b]{Hobbs_2005}. Panel B shows our three-dimensional velocity estimates for the pulsars in their young sample, determined by assuming GC isotropy (Equation \ref{eq3}), in a purple histogram with bins of $50$ km s$^{-1}$. The dashed black line shows our fiducial kick distribution, and the dotted light purple line shows the Maxwellian fit of \citet{Hobbs_2005} with $\sigma=265$ km s$^{-1}$.}
    \label{fig2}
\end{figure*}

For the DM distances (panel F), the fiducial kick distribution lies slightly outside of the confidence interval for the young pulsars, since parallax distance estimates tend to exceed DM distances. However, we are more confident in the parallax distance estimates, which is why our fiducial kick distribution is solely based on the parallax distance estimates. In Appendix \ref{appC}, we compare the distance and velocity estimates based on DM and parallax.

%% file: Sections/4_Maxwellian.tex
\begin{figure*}
    \centering
    \includegraphics[width=18cm]{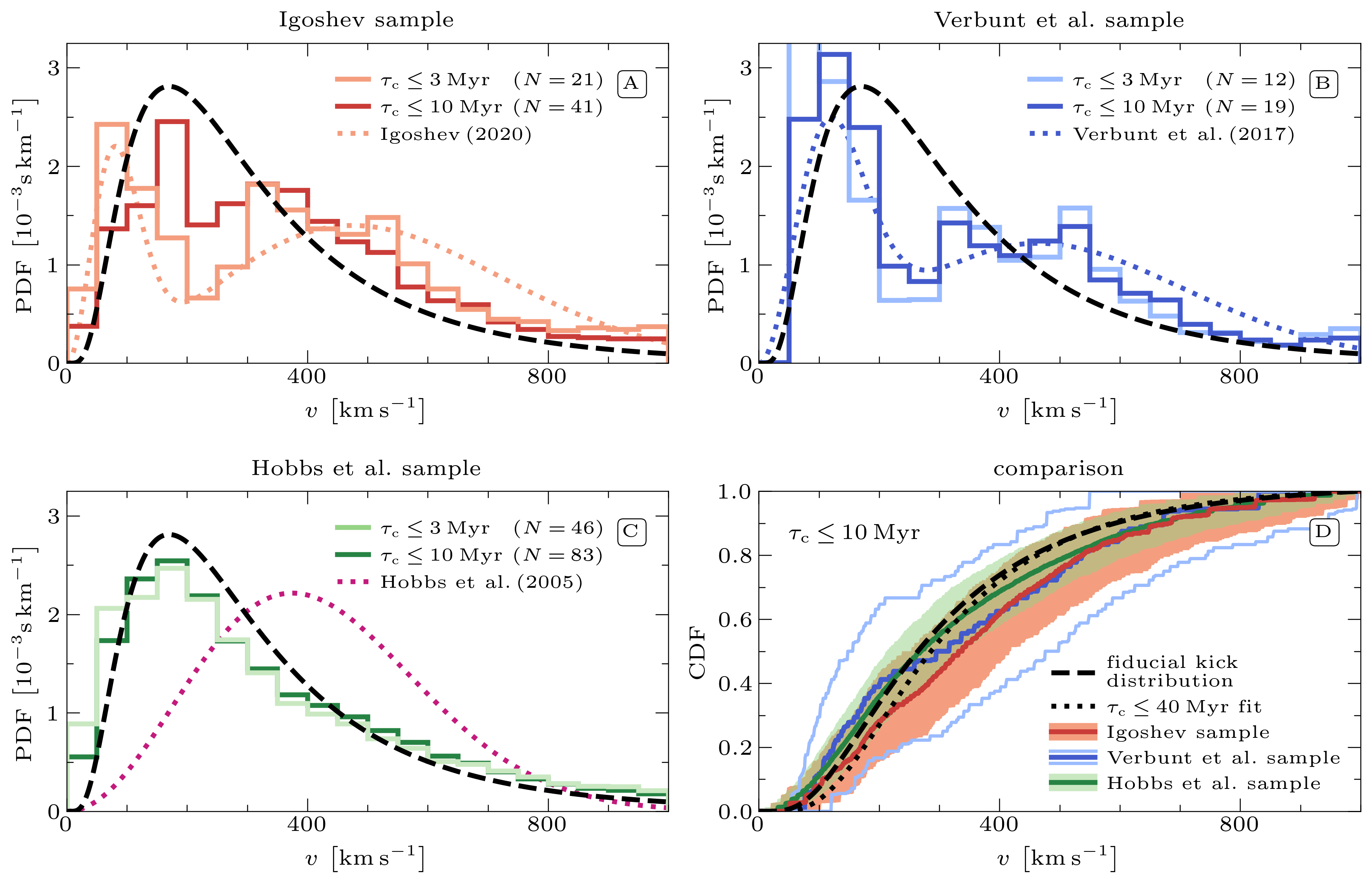}
    \caption{Velocity distributions for the pulsar samples of \citet{Igoshev_2020}, \citet{Verbunt_2017}, and \citet{Hobbs_2005}. Panels A, B, and C show the total velocity distributions for ages below $3$ Myr (light colors) and $10$ Myr (dark colors), in histograms with bins of $50$ km s$^{-1}$. The dashed black lines show our fiducial kick distribution (Equation \ref{eq6}). The dotted light red, light blue, and pink lines show the kick distributions of \citet[][for $\tau_{\text{c}}\leq3$ Myr]{Igoshev_2020}, \citet[][for $\tau_{\text{c}}\leq10$ Myr]{Verbunt_2017}, and \citet[][for $\tau_{\text{c}}\leq3$ Myr]{Hobbs_2005}, respectively. These panels also list the sample sizes (i.e., $N$) for the different samples. Panel D shows median CDFs of the velocity distributions for pulsars with $\tau_{\text{c}}\leq10$ Myr (dark lines), as well as their $95\%$ confidence intervals (light areas and lines), our fiducial kick distribution (black dashed line), and our fit for $\tau_{\text{c}}\leq40$ Myr (black dotted line). The distributions in panel D are normalized between $0$ and $1000$ km s$^{-1}$.}
    \label{fig3}
\end{figure*}

\section{Maxwellian}
\label{sec4}
\noindent Because parallax velocities tend to exceed DM velocities, it is difficult to argue that the fact that \citet{Verbunt_2017} and \citet{Igoshev_2020} find velocities lower than the ones found by \citet{Hobbs_2005} is caused by the difference between parallax and DM distances \citep[as also found by][]{Disberg_2025}. Also, \citet{Valli_2025} find that the kick distribution of \citet{Hobbs_2005} is strongly disfavored for explaining the properties of their Be X-ray binary sample. The velocity estimates of \citet{Hobbs_2005} are based on an analysis of the proper motions and are inferred by applying a \lstinline{CLEAN}-type algorithm \citep{Hogbom_1974} to the estimated transverse velocities of their pulsar sample (for $\tau_{\text{c}}\leq3$ Myr). To their results they fitted a Maxwellian distribution:
\begin{equation}
    \label{eq7}
    p_{_{\text{M}}}(v\text{\hspace{.4mm}}|\text{\hspace{.4mm}}\sigma)=\sqrt{\frac{2}{\pi}}\frac{v^2}{\sigma^3}\exp\left(\frac{-v^2}{2\sigma^2}\right),
\end{equation}
and found that $\sigma=265$ km s$^{-1}$ fits their resulting histogram well. Their Maxwellian fit has a median ${\sim}50\%$ larger than our fiducial kick distribution and thus predicts significantly higher velocities. In an attempt to recreate their results, we consider their values of $v_{t}$ (see their Table 1) which are based on DM distances through the electron density model of \citet{Cordes_2002}, for pulsars with $\tau_{\text{c}}\leq3$ Myr (see their Table 2). Then, we follow \citet{Hobbs_2005} in assuming GC isotropy and estimate the unknown radial velocity by correcting for Solar motion, sampling $10^{3}$ values from Equation \ref{eq3}, and subtracting the velocity of the LSR of the pulsar, in order to sample three-dimensional velocities of the pulsars relative to their LSR.

In panel A of Figure \ref{fig2} we show a histogram of the transverse velocities listed by \citet[][in their Table 1]{Hobbs_2005} in uniform logarithmic bins, as they have done. The height of the histogram bins are not corrected for the logarithmic bin sizes, and since these bin sizes are proportional to $v$ this means that the histogram is effectively showing
\begin{equation}
    \label{eq8}
    p(\log v)=\frac{p(v)\text{d}v}{\text{d}\log v}\propto v\,p(v)\propto\text{bin size}\cdot p(v).
\end{equation}
In other words, this histogram is missing the Jacobian $\text{d}v/\text{d}\log v$ and therefore shows $p(\log v)$ instead of $p(v)$. We find that our histogram of $v_t$ that excludes this Jacobian is identical to the histogram of \citet[][see their Figure 4b]{Hobbs_2005}.

Similarly, if we describe the distribution of three-dimensional velocities with a histogram that does not include a Jacobian and thus shows $p(\log v)$, the result is similar to the histogram of \citet[][see their Figure 7]{Hobbs_2005}. However, to fit the histogrammed empirical distribution to the $p(v)$ Maxwellian of Equation \ref{eq7}, a Jacobian must be included. We replot the histogram in uniform $v$ bins in panel B of Figure \ref{fig2}, including the Jacobian. It is apparent that the empirical distribution contains significantly lower velocities than predicted by the \citet{Hobbs_2005} Maxwellian and is instead well-described by our log-normal fiducial distribution (despite the fact that these velocities are determined assuming GC isotropy instead of LSR isotropy).

Based on our analysis shown in Figure \ref{fig2}, we conclude that the Jacobian that corrects for the logarithmic bin sizes is missing from the analysis of \citet{Hobbs_2005}, meaning that their Maxwellian fit is inconsistent with the data and significantly overestimates the true kick distribution. This explains why \citet{Disberg_2025} are not able to reproduce this Maxwellian, but instead find a distribution more similar to our fiducial kick distribution. The difference between the bimodal kick distributions found by \citet{Verbunt_2017} and \citet{Igoshev_2020} and the Maxwellian of \citet{Hobbs_2005} is therefore not due to inaccuracy in DM distance estimates, but is caused by an erroneous histogram representation due to a missing Jacobian.

%% file: Sections/5_Comparison.tex
\section{Comparison}
\label{sec5}
\noindent In order to compare our results to the velocity distributions found by \citet{Hobbs_2005}, \citet{Verbunt_2017} and \citet{Igoshev_2020}, we plot the histograms of inferred velocities of the young pulsars in their samples (i.e., for $\tau_{\text{c}}\leq3$ Myr as well as $\tau_{\text{c}}\leq10$ Myr) following our methodology described in Section \ref{sec2} (cf.\ Figure \ref{fig1}). In panels A, B, and C of Figure \ref{fig3} we overplot these histograms with the distributions inferred by the respective previous analyses; the histograms for the samples of \citet{Igoshev_2020} and \citet{Verbunt_2017} resemble their bimodal distributions while the \citet{Hobbs_2005} data do not match their Maxwellian because of the Jacobian error. That is, for the Igoshev sample there appears to be a bimodality for $\tau_{\text{c}}\leq3$ Myr and for the Verbunt sample there seems to be a bimodality for both age ranges \citep[as also found by][]{Disberg_2025}. In Appendix \ref{appD} we show that we can reproduce their
double-Maxwellian fits, defined as
\begin{equation}
    \label{eq9}
    p_{_{\text{DM}}}(v\text{\hspace{.4mm}}|\text{\hspace{.4mm}}\sigma_1,\sigma_2,w)=w\,p_{_{\text{M}}}(v\text{\hspace{.4mm}}|\text{\hspace{.4mm}}\sigma_1)+(1-w)\,p_{_{\text{M}}}(v\text{\hspace{.4mm}}|\text{\hspace{.4mm}}\sigma_2),
\end{equation}
where $p_{_{\text{M}}}(v\text{\hspace{.4mm}}|\text{\hspace{.4mm}}\sigma)$ is a Maxwellian distribution (Equation \ref{eq7}), by applying our methodology to their pulsar velocity sample. For the Hobbs sample, we use the Galactic coordinates and proper motions listed in their Table 2 and assume LSR isotropy (Equation \ref{eq4}) as opposed to GC isotropy (Equation \ref{eq3}). We are therefore more confident in these velocity estimates than in the ones shown in Figure \ref{fig2}, and they follow a log-normal distribution well-described by our fiducial kick distribution.

In panel D of Figure \ref{fig3}, we show CDFs of the present-day velocity distributions of these pulsar samples, for $\tau_{\text{c}}\leq10$ Myr. We estimate the $95\%$ confidence intervals by bootstrapping the distributions, similarly to our main results shown in Figure \ref{fig1}. We also show our fiducial kick distribution as well as our fit for $\tau_{\text{c}}\leq40$ Myr, both of which are encompassed by all three confidence intervals. We therefore argue that (1) correcting the results of \citet{Hobbs_2005} for the logarithmic bin sizes reconciles them with the results of \citet{Verbunt_2017} and \citet{Igoshev_2020}, and (2) our fiducial kick distribution is compatible with all of these NS kicks analyses. Nevertheless, we do note that for larger sample sizes we expect the confidence regions of velocity distributions based on DM distances to be slightly below our kick distribution (as shown in Figures \ref{fig1}).

Moreover, despite the fact that there are good reasons why the kick distribution may be bimodal \citep[see e.g.][and references therein]{Katz_1975,Fryer_1998,Podsiadlowski_2004,Fryer_2007,Banerjee_2020,Willcox_2021,Disberg_2025,Valli_2025}, the bimodality found by \citet{Verbunt_2017} and \citet{Igoshev_2020} is not statistically significant: the distributions are well described by our unimodal log-normal fiducial kick distribution. The bimodality is therefore likely a result of low-number statistics. In order to examine this, we randomly select $21$ pulsars (the number of young pulsars used in the \citealt{Igoshev_2020} analysis) from our sample of $\tau_{\text{c}}\leq10$ Myr parallax pulsars and evaluate the histogrammed velocity PDFs. We find that the PDF drops below $1.2\cdot10^{-3}$ s km$^{-1}$ in the 200--250 km s$^{-1}$ in 10\% of the 1000 random draws of $21$ pulsars, showing a bimodality similar to the distributions of \citet{Verbunt_2017} and \citet{Igoshev_2020}. Lastly, we investigate whether a distribution such as a Maxwellian or a (bimodal) double-Maxwellian could provide a better fit to the data than our log-normal model. As shown in Appendix \ref{appE}, we do not find strong evidence in favor of an alternative kick distribution.

%% file: Sections/6_Conclusions.tex
\section{Conclusions}
\label{sec6}
\noindent We have used the proper motions and distances (both based on parallax and DM) of young pulsars in the ATNF Pulsar Catalogue \citep{Manchester_2005} and estimated their present-day velocities relative to their LSR by assuming LSR isotropy (Equation \ref{eq4}). Moreover, we have compared our results to the works of \citet{Hobbs_2005}, \citet{Verbunt_2017}, and \citet{Igoshev_2020}. Based on our findings, we conclude the following:
\begin{itemize}
    \item The present-day velocities of pulsars younger than $10$ Myr, relative to their LSR, are well-described by a log-normal distribution (Equation \ref{eq5}) with $\mu=5.60\pm0.12$ and $\sigma=0.69\pm0.10$, which we adopt as our fiducial kick distribution (Table \ref{tab1} and Figure \ref{fig1}).
    \item The kicks of pulsars older than $10$ Myr, kinematically constrained using the method of \citet{Disberg_2025}, also show a single peak at ${\sim}200$ km s$^{-1}$, similar to the velocities of young pulsars. If we include the inferred kick velocities of pulsars with ages between $10$ Myr and $40$ Myr in our fit, we find a more tightly constrained log-normal distribution with $\mu=5.67\pm0.10$ and $\sigma=0.59\pm0.08$, which is, however, more likely to suffer from systematics.
    \item The velocities based on DM distances are slightly below these results, and therefore cannot explain the difference between the Maxwellian distribution found by \citet{Hobbs_2005} and the double-Maxwellians found by \citet{Verbunt_2017} and \citet{Igoshev_2020}.
    \item \citet{Hobbs_2005} erroneously compared an empirical velocity distribution $p(\log v)$ against a model $p(v)$ without including a Jacobian correction. Therefore, their Maxwellian fit is incorrect and significantly overestimates the true velocities (Figure \ref{fig2}).
    \item Our method can reproduce the double-Maxwellian distributions of \citet{Verbunt_2017} and \citet{Igoshev_2020} relatively accurately. However, we argue that the bimodality that they find is not statistically significant but likely caused by low-number statistics, and our fiducial kick distribution is consistent with their results as well as the results of \citet{Hobbs_2005} that are corrected for the missing Jacobian (Figure \ref{fig3}). 
\end{itemize}
The method we use to estimate the present-day pulsar velocities makes several assumptions. For instance, our fiducial kick distribution does not take into account a likely binary origin of the pulsars, in which the pulsars are kicked out of the binary system \citep[e.g.,][]{Kuranov_2009,Beniamini_2024}. This binary origin can impact the evolutionary history: for example, binary interactions stripping supernova progenitors may affect kick magnitudes \citep{Muller_2018,Willcox_2021}. Also, the pre-supernova orbital velocity of the binary will scatter the speed of the ejected single pulsar around the kick magnitude. Perhaps most importantly, binarity imposes a selection effect: only sufficiently large kicks will unbind binaries, potentially leading to an over-representation of high kicks among apparently single pulsars. However, \citet{Kapil_2023} find that a binary origin does not significantly influence observed kicks in their population-synthesis models. Moreover, we neglect the effects of overdensities in the Galactic disc on the velocities of kicked objects \citep{Ofek_2009} or the potential effect of a \textquotedblleft rocket\textquotedblright-like acceleration due to spin-down radiation \citep{Hirai_2024}. Lastly, we also do not consider a potential observational bias caused by spin-kick alignment \citep{Biryukov_2024}, which may cause pulsar velocities to be overestimated by ${\sim}15\%$ \citep{Mandel_2023}.

%% file: Lay-out/Acknowledgements.tex
\section*{Acknowledgments}
\noindent We thank George Hobbs and Tom Maccarone for useful discussions, and acknowledge
support from the Australian Research Council (ARC) Centre of Excellence for Gravitational-Wave Discovery (OzGrav) through project number CE230100016. In this work, we have used pulsar data from the ATNF Pulsar Catalogue \citep[][available at \href{https://www.atnf.csiro.au/research/pulsar/psrcat/}{https://www.atnf.csiro.au/research/pulsar/psrcat/}]{Manchester_2005}.
\noindent\software{\lstinline{ASTROPY} \citep{Astropy_2013,Astropy_2018,Astropy_2022}, \lstinline{GALPY} \citep{Bovy_2015}, \lstinline{MATPLOTLIB} \citep{Hunter_2007}, \lstinline{NUMPY} \citep{Harris_2020}, \lstinline{SCIPY} \citep{Virtanen_2020}.}

%% file: Lay-out/Appendix.tex
\appendix
\input{Sections/A_Spin_Properties_and_Ages}

\input{Sections/B_Fiducial_Kick_Distribution}
\newpage
\input{Sections/C_Velocities_and_Distances}
\newpage
\input{Sections/D_Maxwellian_Fits}
\newpage
\input{Sections/E_Model_Comparison}

%% file: Sections/A_Spin_Properties_and_Ages.tex
\section{Spin Properties and Ages}
\label{appA}
\noindent The pulsars in our sample (as described in Section \ref{sec2}) have various ages, based on their spin properties. In panel A of Figure \ref{figA}, we show the derivative of the spin period ($\dot{P}$) versus the actual spin period ($P$), which describes pulsar evolutionary stages. There is no significant difference between the distribution of the parallax pulsars and the DM pulsars. We also show an estimate of the \textquotedblleft death line\textquotedblright\ below which pulsars are no longer observable. That is, we use the work of \citet{Beskin_2022} and describe the death line (i.e., $\dot{P}_{\text{dl}}$) as:
\begin{equation}
    \label{eqA1}
    \dot{P}_{\text{dl}}=10^{-15}\beta P^{11/4},
\end{equation}
where they conservatively estimate $\beta=0.01$. The figure shows that only one pulsar in our sample has $\dot{P}<\dot{P}_{\text{dl}}$, so in general our sample agrees with this estimate. Panel B of Figure \ref{figA} shows the distributions of the characteristic ages $\tau_{\text{c}}$ of the pulsars in our sample, where $\gtrsim65\%$ of the pulsars have $\tau_{\text{c}}\leq10$ Myr.
\vfill
\renewcommand{\thefigure}{A}
\begin{figure}[h]
    \centering
    \resizebox{\hsize}{!}{\includegraphics{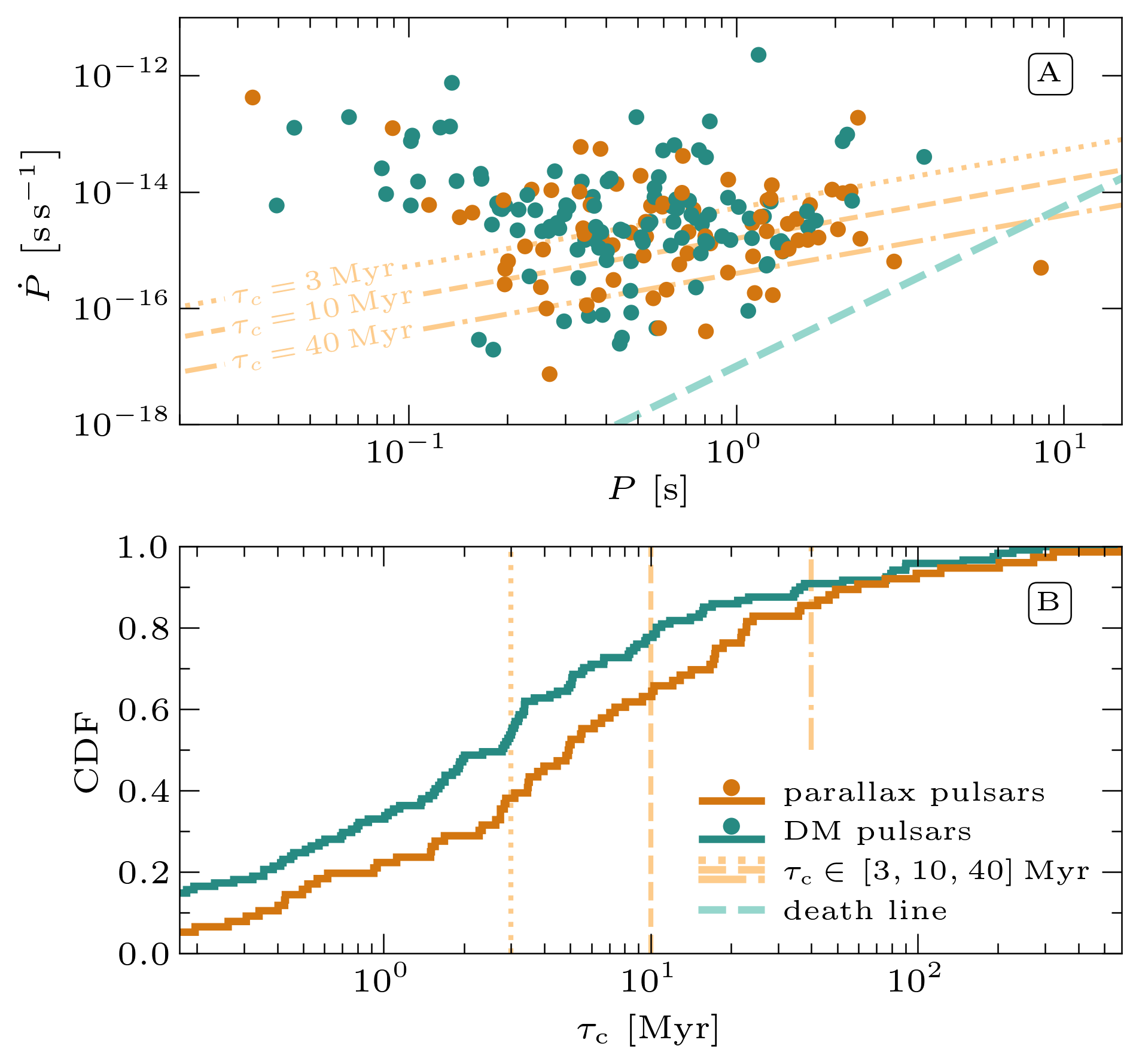}}
    \caption{Spin properties and ages of the pulsars in our sample, for pulsars that have parallax distance estimates (brown) and pulsars that only have a DM distance estimate (teal). Panel A shows the derivative of the spin period ($\dot{P}$) versus the actual spin period ($P$), together with the \textquotedblleft death line\textquotedblright\ (Equation \ref{eqA1}, dashed light teal line). Panel B shows the distributions of the characteristic ages $\tau_{\text{c}}=1/2\cdot P/\dot{P}$. Both panels include lines that indicate values of $\tau_{\text{c}}$ equal to $3$, $10$, and $40$ Myr (light brown lines).}
    \label{figA}
\end{figure}

%% file: Sections/B_Fiducial_Kick_Distribution.tex
\section{Individual Kick Distributions}
\label{appB}
\noindent In Figure \ref{fig1}, we showed histograms containing the most likely pulsar velocities after weighting observations by our fiducial kick distribution. The distributions of $10^{3}$ velocity samples for the individual pulsars are shown in Figure \ref{figB}. We rescaled these distributions by their modes (i.e., $v_{\text{peak}}$), so that they overlap and indicate the general accuracy of the velocity measurements (shown via the $68\%$ intervals). In general, velocity estimates can be expected to be asymmetric since speeds cannot be negative. However, the assumption of isotropy adds an additional asymmetry to these distributions, through the factor of $\cot\theta$ in Equations \ref{eq3} and \ref{eq4}. In Figure \ref{figB}, we show the distribution of $\sqrt{1+\cot^2\theta}-1$, corresponding to a GC isotropic velocity estimate of a 3-d speed based on an observed 2-d velocity with no measurement uncertainty, which contributes significantly to the high velocity tail. Because of this, the total velocity distribution is skewed toward higher velocities.
\vfill
\renewcommand{\thefigure}{B}
\begin{figure}[h]
    \centering
    \resizebox{\hsize}{!}{\includegraphics{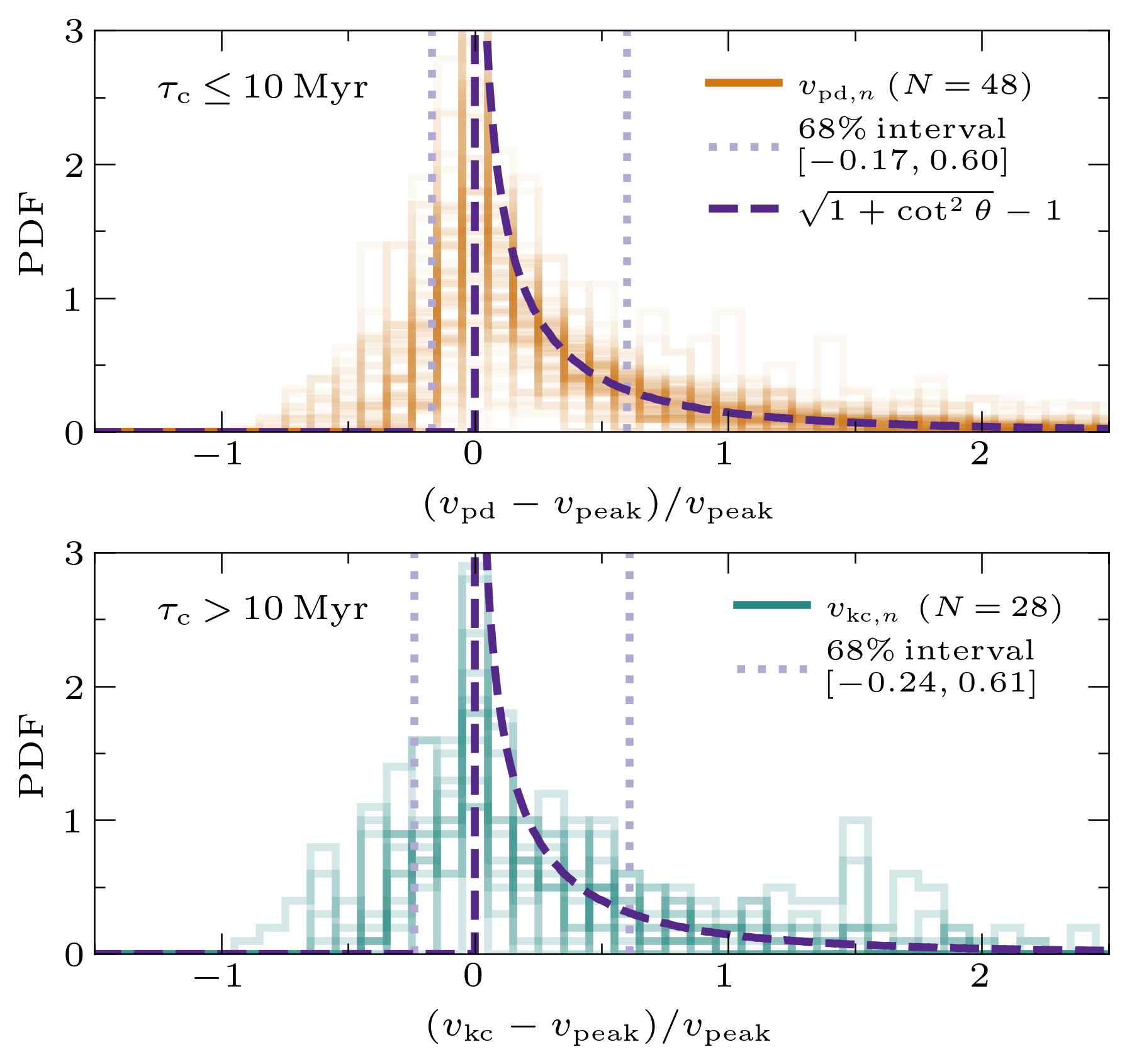}}
    \caption{Individual velocity distributions for each (parallax) pulsar in our sample, for the present-day velocity distributions ($\tau_{\text{c}}\leq10$ Myr, top panel) and the kinematically constrained velocity distributions ($\tau_{\text{c}}>10$ Myr, bottom panel), shown in histograms with bins of $0.1$. The distributions are rescaled based on their mode (i.e., $v_{\text{peak}}$). The dotted light purple lines show the median $68\%$ intervals for these two methods. The dashed purple lines show the impact of the isotropic projection to estimate line-of-sight velocity (see Equations \ref{eq3} and \ref{eq4}).}
    \label{figB}
\end{figure}

%% file: Sections/C_Velocities_and_Distances.tex
\section{Velocities and Distances}
\label{appC}
\noindent Using our method, as described in Section \ref{sec2}, we obtain distributions of parallax distances and DM distances, as well as distributions of the corresponding (transverse and 3-d) velocities. These distributions---displayed in Figure \ref{figC}---take into account the uncertainties on the distances and the isotropy of the (LSR) velocity vector. Panel A shows the distance distributions, where the parallax-based estimates of pulsars typically exceed the DM-based estimates for pulsars with measured parallaxes. However, interestingly, the DM distance distribution of the complete sample aligns more closely with the parallax distance distribution. The differences in the distance distributions translate into similar differences in the transverse and 3-d velocities.
\vfill
\renewcommand{\thefigure}{C}
\begin{figure}[h]
    \centering
    \resizebox{\hsize}{!}{\includegraphics{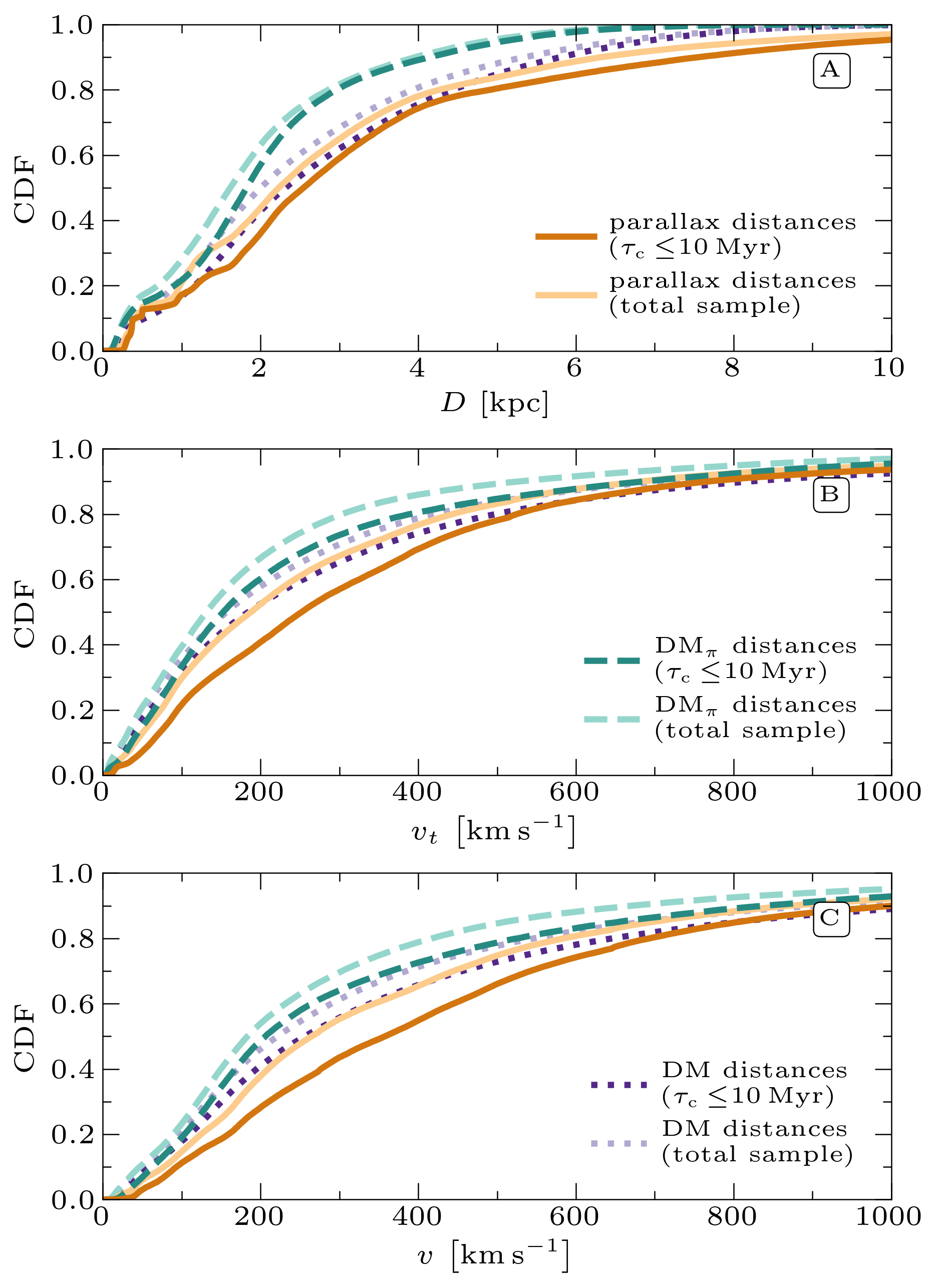}}
    \caption{Distances and velocities of the pulsars in our sample, for parallax distances (brown), DM distances for pulsars with existing parallax estimates (i.e., DM$_{\pi}$, dashed teal), and DM distances for all pulsars (dotted purple), and for both young pulsars (dark) and all ages (light). Panels A, B and C show the distributions of distances, observer-frame transverse velocities, and three-dimensional velocities, respectively.}
    \label{figC}
\end{figure}

%% file: Sections/D_Maxwellian_Fits.tex
\section{Maxwellian Fits}
\label{appD}
\noindent In Figure \ref{fig3}, we show the present-day velocity distributions for the pulsars in the samples of \citet{Igoshev_2020} and \citet{Verbunt_2017}, determined by assuming LSR isotropy (Equation \ref{eq4}) for all objects. The analyses of \citet{Igoshev_2020} and \citet{Verbunt_2017} use a mixed model in which the radial velocities of a fraction of the sample are oriented semi-isotropically, based on the fact that pulsars are thought to be born in the thin disc of the Milky Way and young pulsars are therefore more likely to move away from the disc. Nevertheless, when we fit a double-Maxwellian (Equation \ref{eq9}) to their data processed following our methodology, we find best-fit parameters within the uncertainties of their fits \citep[with the slight exception of $\sigma_{2}$ for the sample of][]{Igoshev_2020}, as shown in Figure \ref{figD}. The main difference is that in our results the contribution of velocities above ${\sim}600$ km s$^{-1}$ is smaller, but this is likely not statistically significant. We therefore conclude that our method relatively accurately reproduces the results of \citet{Igoshev_2020} and \citet{Verbunt_2017}.
\vfill
\renewcommand{\thefigure}{D}
\begin{figure}[h]
    \centering
    \resizebox{\hsize}{!}{\includegraphics{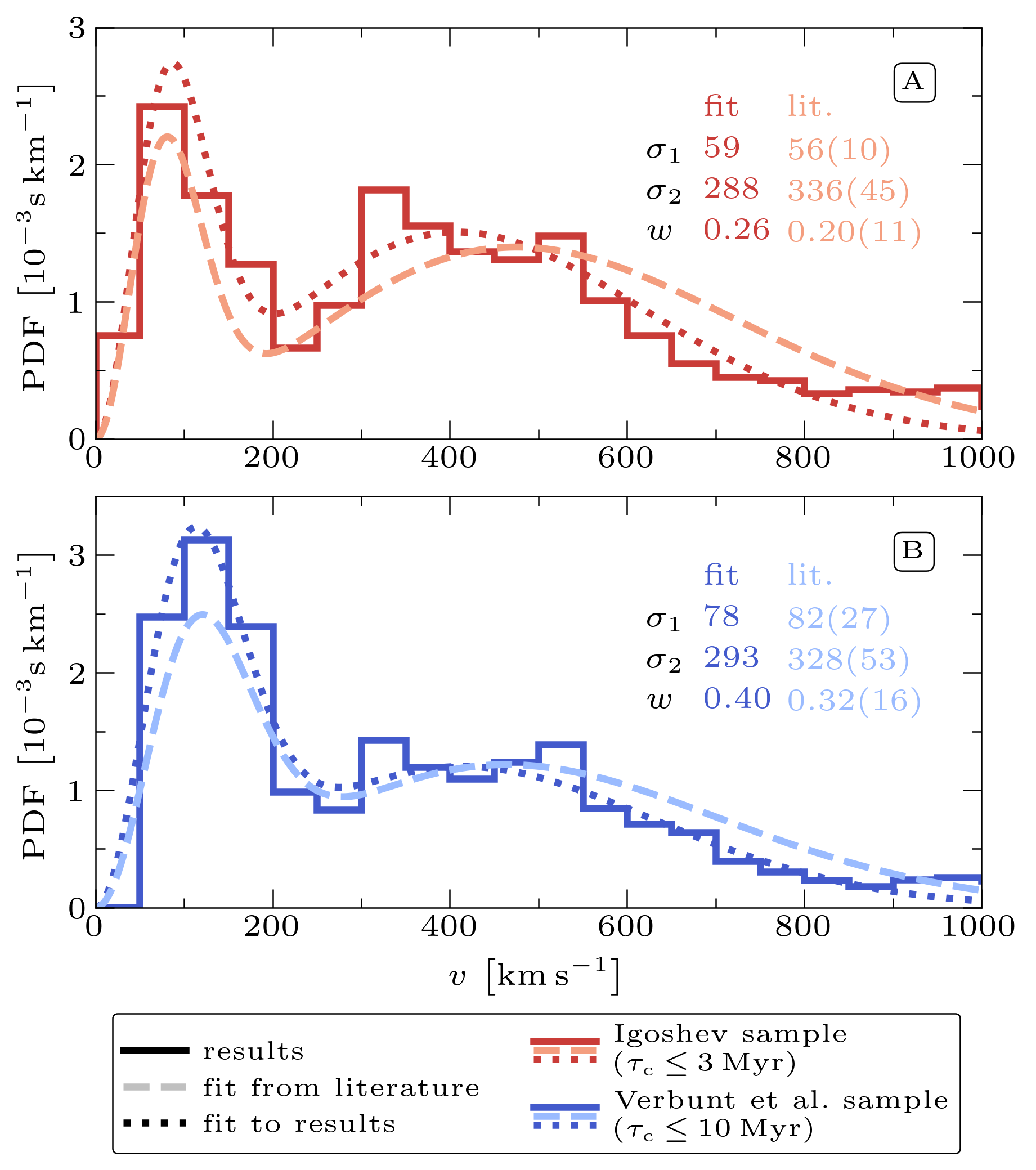}}
    \caption{Velocity distributions for the pulsar samples of \citet{Igoshev_2020} and \citet{Verbunt_2017}, together with their kick distributions (light dashed lines) and our double-Maxwellian fits (dark dotted lines) through Equation \ref{eq9}. The parameters of our fits and those from the literature are listed in the figure as well, along with quoted uncertainties on the latter in parentheses.}
    \label{figD}
\end{figure}

%% file: Sections/E_Model_Comparison.tex
\section{Model Comparison}
\label{appE}
\noindent Our fiducial kick distribution is a log-normal distribution fitted to the pulsar velocity estimates from our sample, and in Figure \ref{fig1} we show that this fit stays within the bootstrapped $95\%$ confidence region of the observed CDFs. However, the question arises whether an alternative model, such as a double-Maxwellian model, might be a better description of our results. We therefore fit several distributions to our results for the $48$ parallax pulsars with $\tau_{\text{c}}\leq10$ Myr: a Maxwellian, a double-Maxwellian, and a Weibull distribution---a unimodal distribution defined as:
\renewcommand{\theequation}{E1}
\begin{equation}
    \label{eqE1}
    p_{_{\text{W}}}(v\text{\hspace{.4mm}}|\text{\hspace{.4mm}}\lambda,k)=\dfrac{k}{\lambda}\left(\dfrac{v}{\lambda}\right)^{k-1}\exp\left(-\left(\dfrac{v}{\lambda}\right)^k\right),
\end{equation}
where the parameters $k$ and $\lambda$ determine the shape and scale of the distribution, respectively.

For these fitted distributions, as well as our log-normal fiducial kick distribution and the literature distributions of \citet{Hobbs_2005}, \citet{Verbunt_2017} and \citet{Igoshev_2020}, we evaluate the Bayesian information criterion (BIC), defined as: 
\renewcommand{\theequation}{E2}
\begin{equation}
    \label{eqE2}
    \text{BIC}=P\ln N-2\ln\hat{\mathcal{L}},
\end{equation}
where $P$ equals the number of parameters in the distribution function, $N$ is the number of pulsars in the sample, and $\hat{\mathcal{L}}$ is the maximum likelihood (Equation \ref{eq6}). \citet{Kass_1995} state that a difference in BIC between $6$ and $10$ corresponds to \textquotedblleft strong\textquotedblright\ evidence in favor of the model with the lowest BIC, and if the difference exceeds $10$ the evidence is \textquotedblleft decisive\textquotedblright.

\renewcommand{\thefootnote}{3}
In Table \ref{tabE}, we list the fitted parameters of the considered distributions, as well as the values of the BIC and the difference between their BIC and the BIC of our fiducial kick distribution ($\Delta$BIC). These distributions are shown in Figure \ref{figE}. For the fitted Maxwellian, double-Maxwellian, and Weibull distribution, the value of $|\Delta\text{BIC}|$ is less than $4$, meaning there is no strong evidence in favor of either distribution. We find that $N\gtrsim30$ hypothetical pulsars with no measurement uncertainty with kicks sampled from the \citet{Igoshev_2020} distribution would be sufficient to obtain $|\Delta\text{BIC}|>10$ between the fitted log-normal and double-Maxwellian distributions, meaning that the small $\Delta$BIC values are mainly caused by observational uncertainties and the necessity to estimate $v_{r}$ (Equation \ref{eq4}). Although the Maxwellian has a significantly lower likelihood, it has the benefit of having only one parameter. Moreover, the best-fit double-Maxwellian to our larger sample is significantly less bimodal than the distributions shown in Figure \ref{figD}. We note that if we fit a scaled beta distribution \citep{ODoherty_2023} or an asymmetric Gaussian \citep{Disberg_2023}, the resulting BIC values are also within the range where these functional forms are neither preferred nor ruled out given the available data. Lastly, the literature distributions, taking the best-fit values from the literature,\footnote{We still account for the full number of parameters $P$ for the relevant distributions when computing BICs given the largely overlapping data sets, though one could arguably penalize these distributions less given that there are no free parameters in the quoted fits.} have higher BIC values, where there is strong evidence against the (erroneous) distribution of \citet{Hobbs_2005} and decisive evidence against the distribution of \citet{Igoshev_2020}.
\textcolor{white}{.\\ .\\ .} 
\input{Lay-out/TableE}
\renewcommand{\thefigure}{E}
\begin{figure}
    \centering
    \resizebox{\hsize}{!}{\includegraphics{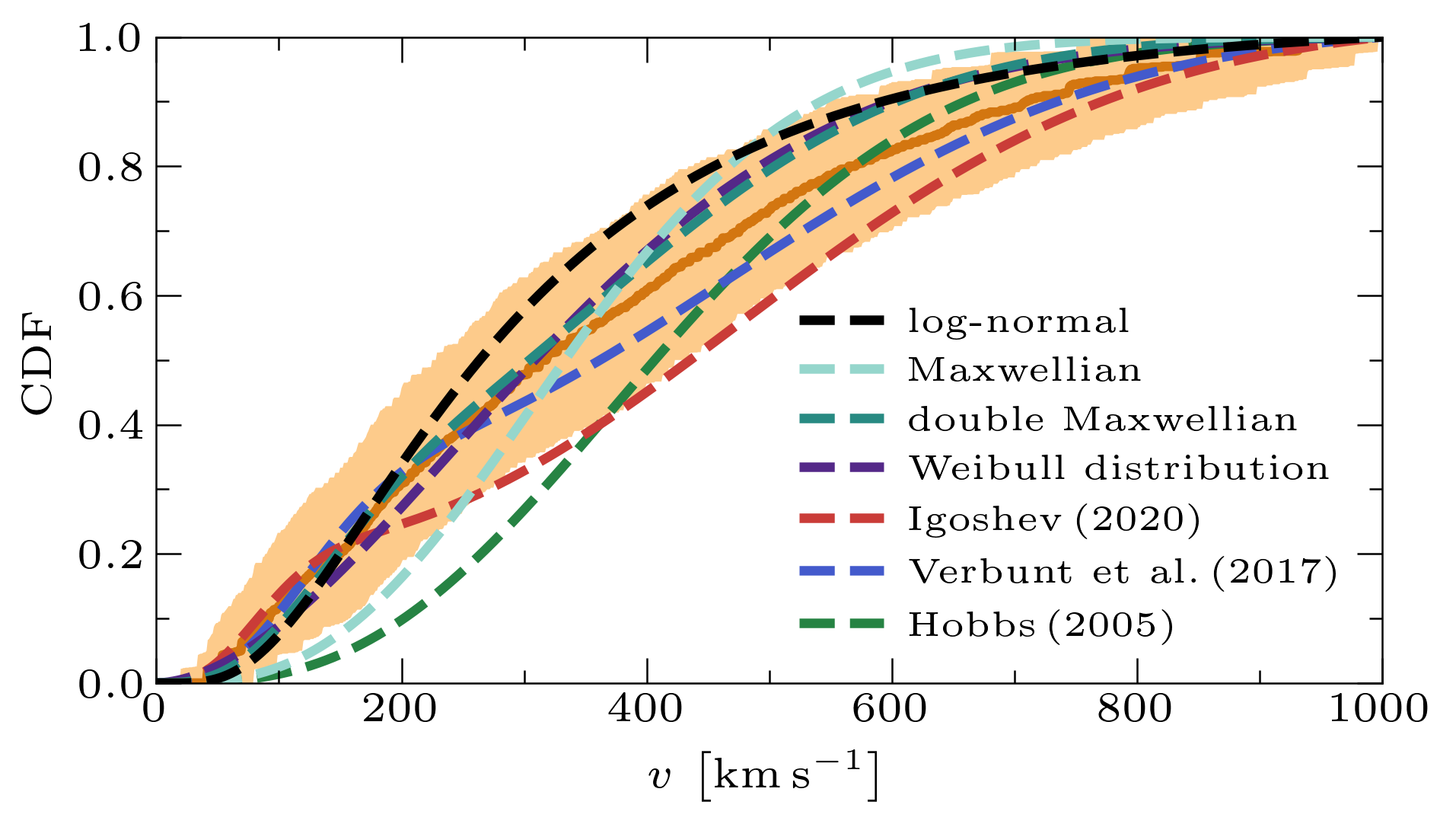}}
    \caption{CDFs of the distributions fitted to our data set: our log-normal fiducial kick distribution (Equation \ref{eq5}, black), a Maxwellian (Equation \ref{eq7}, light teal), a double-Maxwellian (Equation \ref{eq9}, teal), and a Weibull distribution (Equation \ref{eqE1}, purple), together with the distributions of \citet[][red]{Igoshev_2020}, \citet[][blue]{Verbunt_2017} and \citet[][green]{Hobbs_2005}. The orange line is our median bootstrapped distribution and the light orange region corresponds to the $95\%$ confidence region (Figure \ref{fig1}). All distributions are normalized between $0$ and $1000$ km s$^{-1}$.}
    \label{figE}
\end{figure}

%% file: Lay-out/TableE.tex
\renewcommand{\thetable}{E}
\begin{table}
\caption{BIC values (Equation \ref{eqE2}) of (fitted) distributions.\label{tabE}}
\hspace{-10mm}\begin{tabular}{llcc}
\hline\hline\\[-13pt]
distribution & fit & BIC & $\Delta$BIC\\
[2pt]\hline\\[-13pt]
log-normal & & 666.2 & 0\\
Maxwellian & $\sigma=217(17)$ & 666.4 & 0.2\\
double-Maxwellian & $\sigma_1=90(50)$ & 665.7 & -0.5\\
 & $\sigma_2=260(40)$ &\\
 & $w=0.3(2)$ & \\
Weibull distribution & $\lambda=380(40)$ & 662.7 & -3.5\\
 & $k=1.8(3)$ & &\\
\citet{Hobbs_2005} & & 673.1 & 6.9\\
\citet{Verbunt_2017} & & 671.3 & 5.1\\
\citet{Igoshev_2020} & & 678.2 & 12\\
[2pt]\hline
\end{tabular}
\tablecomments{\footnotesize The values in parentheses are half the widths of the $68\%$ credible intervals. The log-normal distribution is our fiducial kick distribution (Table \ref{tab1}). The value of $\Delta$BIC equals the BIC of the distribution minus the BIC of our fiducial kick distribution. The parameters $\sigma$, $\sigma_1$, $\sigma_2$, and $\lambda$ are given in km s$^{-1}$.}
\end{table}

%% file: Lay-out/References.tex
\bibliography{references}{}
\bibliographystyle{aas_v6}

%% file: main.bbl
\begin{thebibliography}{}
\expandafter\ifx\csname natexlab\endcsname\relax\def\natexlab#1{#1}\fi
\providecommand{\url}[1]{\href{#1}{#1}}
\providecommand{\dodoi}[1]{doi:~\href{http://doi.org/#1}{\nolinkurl{#1}}}
\providecommand{\doeprint}[1]{\href{http://ascl.net/#1}{\nolinkurl{http://ascl.net/#1}}}
\providecommand{\doarXiv}[1]{\href{https://arxiv.org/abs/#1}{\nolinkurl{https://arxiv.org/abs/#1}}}

\bibitem[{{Amend} {et~al.}(2024){Amend}, {Fryer}, {Mumpower}, \& {Korobkin}}]{Amend_2024}
{Amend}, B., {Fryer}, C.~L., {Mumpower}, M.~R., \& {Korobkin}, O. 2024, arXiv e-prints, arXiv:2412.05424, \dodoi{10.48550/arXiv.2412.05424}

\bibitem[{{Arzoumanian} {et~al.}(2002){Arzoumanian}, {Chernoff}, \& {Cordes}}]{Arzoumanian_2002}
{Arzoumanian}, Z., {Chernoff}, D.~F., \& {Cordes}, J.~M. 2002, \apj, 568, 289, \dodoi{10.1086/338805}

\bibitem[{{Astropy Collaboration} {et~al.}(2013)}]{Astropy_2013}
{Astropy Collaboration}, {et~al.} 2013, \aap, 558, A33, \dodoi{10.1051/0004-6361/201322068}

\bibitem[{{Astropy Collaboration} {et~al.}(2018)}]{Astropy_2018}
---. 2018, \aj, 156, 123, \dodoi{10.3847/1538-3881/aabc4f}

\bibitem[{{Astropy Collaboration} {et~al.}(2022)}]{Astropy_2022}
---. 2022, \apj, 935, 167, \dodoi{10.3847/1538-4357/ac7c74}

\bibitem[{{Atri} {et~al.}(2019){Atri}, {Miller-Jones}, {Bahramian}, {Plotkin}, {Jonker}, {Nelemans}, {Maccarone}, {Sivakoff}, {Deller}, {Chaty}, {Torres}, {Horiuchi}, {McCallum}, {Natusch}, {Phillips}, {Stevens}, \& {Weston}}]{Atri_2019}
{Atri}, P., {Miller-Jones}, J.~C.~A., {Bahramian}, A., {et~al.} 2019, \mnras, 489, 3116, \dodoi{10.1093/mnras/stz2335}

\bibitem[{{Banerjee} {et~al.}(2020){Banerjee}, {Belczynski}, {Fryer}, {Berczik}, {Hurley}, {Spurzem}, \& {Wang}}]{Banerjee_2020}
{Banerjee}, S., {Belczynski}, K., {Fryer}, C.~L., {et~al.} 2020, \aap, 639, A41, \dodoi{10.1051/0004-6361/201935332}

\bibitem[{{Beniamini} \& {Piran}(2024)}]{Beniamini_2024}
{Beniamini}, P., \& {Piran}, T. 2024, \apj, 966, 17, \dodoi{10.3847/1538-4357/ad32cd}

\bibitem[{{Beskin} \& {Istomin}(2022)}]{Beskin_2022}
{Beskin}, V.~S., \& {Istomin}, A.~Y. 2022, \mnras, 516, 5084, \dodoi{10.1093/mnras/stac2423}

\bibitem[{{Biryukov} \& {Beskin}(2024)}]{Biryukov_2024}
{Biryukov}, A., \& {Beskin}, G. 2024, arXiv e-prints, arXiv:2412.12017, \dodoi{10.48550/arXiv.2412.12017}

\bibitem[{{Bovy}(2015)}]{Bovy_2015}
{Bovy}, J. 2015, \apjs, 216, 29, \dodoi{10.1088/0067-0049/216/2/29}

\bibitem[{{Brisken} {et~al.}(2002){Brisken}, {Benson}, {Goss}, \& {Thorsett}}]{Brisken_2002}
{Brisken}, W.~F., {Benson}, J.~M., {Goss}, W.~M., \& {Thorsett}, S.~E. 2002, \apj, 571, 906, \dodoi{10.1086/340098}

\bibitem[{{Brisken} {et~al.}(2003){Brisken}, {Thorsett}, {Golden}, \& {Goss}}]{Brisken_2003}
{Brisken}, W.~F., {Thorsett}, S.~E., {Golden}, A., \& {Goss}, W.~M. 2003, \apjl, 593, L89, \dodoi{10.1086/378184}

\bibitem[{{Bruzewski} {et~al.}(2023){Bruzewski}, {Schinzel}, {Taylor}, {Demorest}, {Frail}, {Kerr}, \& {Kumar}}]{Bruzewski_2023}
{Bruzewski}, S., {Schinzel}, F.~K., {Taylor}, G.~B., {et~al.} 2023, \apj, 958, 163, \dodoi{10.3847/1538-4357/ad07e4}

\bibitem[{{Burrows} {et~al.}(1995){Burrows}, {Hayes}, \& {Fryxell}}]{Burrows_1995}
{Burrows}, A., {Hayes}, J., \& {Fryxell}, B.~A. 1995, \apj, 450, 830, \dodoi{10.1086/176188}

\bibitem[{{Burrows} {et~al.}(2024){Burrows}, {Wang}, {Vartanyan}, \& {Coleman}}]{Burrows_2024}
{Burrows}, A., {Wang}, T., {Vartanyan}, D., \& {Coleman}, M. S.~B. 2024, \apj, 963, 63, \dodoi{10.3847/1538-4357/ad2353}

\bibitem[{{Chatterjee} {et~al.}(2001){Chatterjee}, {Cordes}, {Lazio}, {Goss}, {Fomalont}, \& {Benson}}]{Chatterjee_2001}
{Chatterjee}, S., {Cordes}, J.~M., {Lazio}, T.~J.~W., {et~al.} 2001, \apj, 550, 287, \dodoi{10.1086/319735}

\bibitem[{{Chatterjee} {et~al.}(2004){Chatterjee}, {Cordes}, {Vlemmings}, {Arzoumanian}, {Goss}, \& {Lazio}}]{Chatterjee_2004}
{Chatterjee}, S., {Cordes}, J.~M., {Vlemmings}, W.~H.~T., {et~al.} 2004, \apj, 604, 339, \dodoi{10.1086/381748}

\bibitem[{{Chatterjee} {et~al.}(2009){Chatterjee}, {Brisken}, {Vlemmings}, {Goss}, {Lazio}, {Cordes}, {Thorsett}, {Fomalont}, {Lyne}, \& {Kramer}}]{Chatterjee_2009}
{Chatterjee}, S., {Brisken}, W.~F., {Vlemmings}, W.~H.~T., {et~al.} 2009, \apj, 698, 250, \dodoi{10.1088/0004-637X/698/1/250}

\bibitem[{{Cordes} \& {Chernoff}(1998)}]{Cordes_1998}
{Cordes}, J.~M., \& {Chernoff}, D.~F. 1998, \apj, 505, 315, \dodoi{10.1086/306138}

\bibitem[{{Cordes} \& {Lazio}(2002)}]{Cordes_2002}
{Cordes}, J.~M., \& {Lazio}, T.~J.~W. 2002, arXiv e-prints, astro, \dodoi{10.48550/arXiv.astro-ph/0207156}

\bibitem[{{Deller} {et~al.}(2009){Deller}, {Tingay}, {Bailes}, \& {Reynolds}}]{Deller_2009}
{Deller}, A.~T., {Tingay}, S.~J., {Bailes}, M., \& {Reynolds}, J.~E. 2009, \apj, 701, 1243, \dodoi{10.1088/0004-637X/701/2/1243}

\bibitem[{{Deller} {et~al.}(2019){Deller}, {Goss}, {Brisken}, {Chatterjee}, {Cordes}, {Janssen}, {Kovalev}, {Lazio}, {Petrov}, {Stappers}, \& {Lyne}}]{Deller_2019}
{Deller}, A.~T., {Goss}, W.~M., {Brisken}, W.~F., {et~al.} 2019, \apj, 875, 100, \dodoi{10.3847/1538-4357/ab11c7}

\bibitem[{{Ding} {et~al.}(2024){Ding}, {Deller}, {Swiggum}, {Lynch}, {Chatterjee}, \& {Tauris}}]{Ding_2024}
{Ding}, H., {Deller}, A.~T., {Swiggum}, J.~K., {et~al.} 2024, \apj, 970, 90, \dodoi{10.3847/1538-4357/ad4883}

\bibitem[{{Disberg} {et~al.}(2024{\natexlab{a}}){Disberg}, {Gaspari}, \& {Levan}}]{Disberg_2024a}
{Disberg}, P., {Gaspari}, N., \& {Levan}, A.~J. 2024{\natexlab{a}}, \aap, 687, A272, \dodoi{10.1051/0004-6361/202449996}

\bibitem[{{Disberg} {et~al.}(2024{\natexlab{b}}){Disberg}, {Gaspari}, \& {Levan}}]{Disberg_2024b}
---. 2024{\natexlab{b}}, \aap, 689, A348, \dodoi{10.1051/0004-6361/202450790}

\bibitem[{{Disberg} {et~al.}(2025){Disberg}, {Gaspari}, \& {Levan}}]{Disberg_2025}
---. 2025, arXiv e-prints, arXiv:2503.01429, \dodoi{10.48550/arXiv.2503.01429}

\bibitem[{{Disberg} \& {Nelemans}(2023)}]{Disberg_2023}
{Disberg}, P., \& {Nelemans}, G. 2023, \aap, 676, A31, \dodoi{10.1051/0004-6361/202245693}

\bibitem[{{Faucher-Gigu{\`e}re} \& {Kaspi}(2006)}]{Faucher_2006}
{Faucher-Gigu{\`e}re}, C.-A., \& {Kaspi}, V.~M. 2006, \apj, 643, 332, \dodoi{10.1086/501516}

\bibitem[{{Fortin} {et~al.}(2022){Fortin}, {Garc{\'\i}a}, {Chaty}, {Chassande-Mottin}, \& {Simaz Bunzel}}]{Fortin_2022}
{Fortin}, F., {Garc{\'\i}a}, F., {Chaty}, S., {Chassande-Mottin}, E., \& {Simaz Bunzel}, A. 2022, \aap, 665, A31, \dodoi{10.1051/0004-6361/202140853}

\bibitem[{{Fryer} {et~al.}(1998){Fryer}, {Burrows}, \& {Benz}}]{Fryer_1998}
{Fryer}, C., {Burrows}, A., \& {Benz}, W. 1998, \apj, 496, 333, \dodoi{10.1086/305348}

\bibitem[{{Fryer} \& {Young}(2007)}]{Fryer_2007}
{Fryer}, C.~L., \& {Young}, P.~A. 2007, \apj, 659, 1438, \dodoi{10.1086/513003}

\bibitem[{{Gaspari} {et~al.}(2024){Gaspari}, {Levan}, {Chrimes}, \& {Nelemans}}]{Gaspari_2024a}
{Gaspari}, N., {Levan}, A.~J., {Chrimes}, A.~A., \& {Nelemans}, G. 2024, \mnras, 527, 1101, \dodoi{10.1093/mnras/stad3259}

\bibitem[{{GRAVITY Collab.} {et~al.}(2018){GRAVITY Collab.}, {Abuter}, {Amorim}, {Anugu}, {Baub{\"o}ck}, {Benisty}, {Berger}, {Blind}, {Bonnet}, {Brandner}, {Buron}, {Collin}, {Chapron}, {Cl{\'e}net}, {Coud{\'e} Du Foresto}, {de Zeeuw}, {Deen}, {Delplancke-Str{\"o}bele}, {Dembet}, {Dexter}, {Duvert}, {Eckart}, {Eisenhauer}, {Finger}, {F{\"o}rster Schreiber}, {F{\'e}dou}, {Garcia}, {Garcia Lopez}, {Gao}, {Gendron}, {Genzel}, {Gillessen}, {Gordo}, {Habibi}, {Haubois}, {Haug}, {Hau{\ss}mann}, {Henning}, {Hippler}, {Horrobin}, {Hubert}, {Hubin}, {Jimenez Rosales}, {Jochum}, {Jocou}, {Kaufer}, {Kellner}, {Kendrew}, {Kervella}, {Kok}, {Kulas}, {Lacour}, {Lapeyr{\`e}re}, {Lazareff}, {Le Bouquin}, {L{\'e}na}, {Lippa}, {Lenzen}, {M{\'e}rand}, {M{\"u}ler}, {Neumann}, {Ott}, {Palanca}, {Paumard}, {Pasquini}, {Perraut}, {Perrin}, {Pfuhl}, {Plewa}, {Rabien}, {Ram{\'\i}rez}, {Ramos}, {Rau}, {Rodr{\'\i}guez-Coira}, {Rohloff}, {Rousset}, {Sanchez-Bermudez}, {Scheithauer}, {Sch{\"o}ller}, {Schuler}, {Spyromilio}, {Straub},
  {Straubmeier}, {Sturm}, {Tacconi}, {Tristram}, {Vincent}, {von Fellenberg}, {Wank}, {Waisberg}, {Widmann}, {Wieprecht}, {Wiest}, {Wiezorrek}, {Woillez}, {Yazici}, {Ziegler}, \& {Zins}}]{Gravity_2018}
{GRAVITY Collab.}, {Abuter}, R., {Amorim}, A., {et~al.} 2018, \aap, 615, L15, \dodoi{10.1051/0004-6361/201833718}

\bibitem[{{Gunn} \& {Ostriker}(1970)}]{Gunn_1970}
{Gunn}, J.~E., \& {Ostriker}, J.~P. 1970, \apj, 160, 979, \dodoi{10.1086/150487}

\bibitem[{{Hansen} \& {Phinney}(1997)}]{Hansen_1997}
{Hansen}, B. M.~S., \& {Phinney}, E.~S. 1997, \mnras, 291, 569, \dodoi{10.1093/mnras/291.3.569}

\bibitem[{{Harris} {et~al.}(2020){Harris}, {Millman}, {Van der Walt}, {Gommers}, {Virtanen}, {Cournapeau}, {Wieser}, {Taylor}, {Berg}, {Smith}, {Kern}, {Picus}, {Hoyer}, {Van Kerkwijk}, {Brett}, {Haldane}, {del R{\'\i}o}, {Wiebe}, {Peterson}, {G{\'e}rard-Marchant}, {Sheppard}, {Reddy}, {Weckesser}, {Abbasi}, {Gohlke}, \& {Oliphant}}]{Harris_2020}
{Harris}, C.~R., {Millman}, K.~J., {Van der Walt}, S.~J., {et~al.} 2020, \nat, 585, 357, \dodoi{10.1038/s41586-020-2649-2}

\bibitem[{{Herant}(1995)}]{Herant_1995}
{Herant}, M. 1995, \physrep, 256, 117, \dodoi{10.1016/0370-1573(94)00105-C}

\bibitem[{{Hirai} {et~al.}(2024){Hirai}, {Podsiadlowski}, {Heger}, \& {Nagakura}}]{Hirai_2024}
{Hirai}, R., {Podsiadlowski}, P., {Heger}, A., \& {Nagakura}, H. 2024, \apjl, 972, L18, \dodoi{10.3847/2041-8213/ad6e77}

\bibitem[{{Hobbs} {et~al.}(2005){Hobbs}, {Lorimer}, {Lyne}, \& {Kramer}}]{Hobbs_2005}
{Hobbs}, G., {Lorimer}, D.~R., {Lyne}, A.~G., \& {Kramer}, M. 2005, \mnras, 360, 974, \dodoi{10.1111/j.1365-2966.2005.09087.x}

\bibitem[{{H{\"o}gbom}(1974)}]{Hogbom_1974}
{H{\"o}gbom}, J.~A. 1974, \aaps, 15, 417

\bibitem[{{Hunter}(2007)}]{Hunter_2007}
{Hunter}, J.~D. 2007, Computing in Science and Engineering, 9, 90, \dodoi{10.1109/MCSE.2007.55}

\bibitem[{{Igoshev}(2019)}]{Igoshev_2019}
{Igoshev}, A.~P. 2019, \mnras, 482, 3415, \dodoi{10.1093/mnras/sty2945}

\bibitem[{{Igoshev}(2020)}]{Igoshev_2020}
---. 2020, \mnras, 494, 3663, \dodoi{10.1093/mnras/staa958}

\bibitem[{{Igoshev} {et~al.}(2021){Igoshev}, {Chruslinska}, {Dorozsmai}, \& {Toonen}}]{Igoshev_2021}
{Igoshev}, A.~P., {Chruslinska}, M., {Dorozsmai}, A., \& {Toonen}, S. 2021, \mnras, 508, 3345, \dodoi{10.1093/mnras/stab2734}

\bibitem[{{Igoshev} {et~al.}(2016){Igoshev}, {Verbunt}, \& {Cator}}]{Igoshev_2016}
{Igoshev}, A.~P., {Verbunt}, F., \& {Cator}, E. 2016, \aap, 591, A123, \dodoi{10.1051/0004-6361/201527471}

\bibitem[{{Janka} \& {Müller}(1994)}]{Janka_1994}
{Janka}, H.~T., \& {Müller}, E. 1994, \aap, 290, 496

\bibitem[{{Kapil} {et~al.}(2023){Kapil}, {Mandel}, {Berti}, \& {M{\"u}ller}}]{Kapil_2023}
{Kapil}, V., {Mandel}, I., {Berti}, E., \& {M{\"u}ller}, B. 2023, \mnras, 519, 5893, \dodoi{10.1093/mnras/stad019}

\bibitem[{Kass \& Raftery(1995)}]{Kass_1995}
Kass, R.~E., \& Raftery, A.~E. 1995, Journal of the American Statistical Association, 90, 773.
\newblock \url{http://www.jstor.org/stable/2291091}

\bibitem[{{Katz}(1975)}]{Katz_1975}
{Katz}, J.~I. 1975, \nat, 253, 698, \dodoi{10.1038/253698a0}

\bibitem[{{Keith} {et~al.}(2024){Keith}, {Johnston}, {Karastergiou}, {Weltevrede}, {Lower}, {Basu}, {Posselt}, {Oswald}, {Parthasarathy}, {Cameron}, {Serylak}, \& {Buchner}}]{Keith_2024}
{Keith}, M.~J., {Johnston}, S., {Karastergiou}, A., {et~al.} 2024, \mnras, 530, 1581, \dodoi{10.1093/mnras/stae937}

\bibitem[{{Kirsten} {et~al.}(2015){Kirsten}, {Vlemmings}, {Campbell}, {Kramer}, \& {Chatterjee}}]{Kirsten_2015}
{Kirsten}, F., {Vlemmings}, W., {Campbell}, R.~M., {Kramer}, M., \& {Chatterjee}, S. 2015, \aap, 577, A111, \dodoi{10.1051/0004-6361/201425562}

\bibitem[{{Kuranov} {et~al.}(2009){Kuranov}, {Popov}, \& {Postnov}}]{Kuranov_2009}
{Kuranov}, A.~G., {Popov}, S.~B., \& {Postnov}, K.~A. 2009, \mnras, 395, 2087, \dodoi{10.1111/j.1365-2966.2009.14595.x}

\bibitem[{{Lin} {et~al.}(2023){Lin}, {Van Kerkwijk}, {Kirsten}, {Pen}, \& {Deller}}]{Lin_2023}
{Lin}, R., {Van Kerkwijk}, M.~H., {Kirsten}, F., {Pen}, U.-L., \& {Deller}, A.~T. 2023, \apj, 952, 161, \dodoi{10.3847/1538-4357/acdc98}

\bibitem[{{Lorimer} {et~al.}(2006){Lorimer}, {Faulkner}, {Lyne}, {Manchester}, {Kramer}, {McLaughlin}, {Hobbs}, {Possenti}, {Stairs}, {Camilo}, {Burgay}, {D'Amico}, {Corongiu}, \& {Crawford}}]{Lorimer_2006}
{Lorimer}, D.~R., {Faulkner}, A.~J., {Lyne}, A.~G., {et~al.} 2006, \mnras, 372, 777, \dodoi{10.1111/j.1365-2966.2006.10887.x}

\bibitem[{{Lyne} \& {Lorimer}(1994)}]{Lyne_1994}
{Lyne}, A.~G., \& {Lorimer}, D.~R. 1994, \nat, 369, 127, \dodoi{10.1038/369127a0}

\bibitem[{{Lyne} {et~al.}(1996){Lyne}, {Pritchard}, {Graham-Smith}, \& {Camilo}}]{Lyne_1996}
{Lyne}, A.~G., {Pritchard}, R.~S., {Graham-Smith}, F., \& {Camilo}, F. 1996, \nat, 381, 497, \dodoi{10.1038/381497a0}

\bibitem[{{Manchester} {et~al.}(2005){Manchester}, {Hobbs}, {Teoh}, \& {Hobbs}}]{Manchester_2005}
{Manchester}, R.~N., {Hobbs}, G.~B., {Teoh}, A., \& {Hobbs}, M. 2005, \aj, 129, 1993, \dodoi{10.1086/428488}

\bibitem[{{Mandel} {et~al.}(2019){Mandel}, {Farr}, \& {Gair}}]{Mandel_2019}
{Mandel}, I., {Farr}, W.~M., \& {Gair}, J.~R. 2019, \mnras, 486, 1086, \dodoi{10.1093/mnras/stz896}

\bibitem[{{Mandel} \& {Igoshev}(2023)}]{Mandel_2023}
{Mandel}, I., \& {Igoshev}, A.~P. 2023, \apj, 944, 153, \dodoi{10.3847/1538-4357/acb3c3}

\bibitem[{{Maoz} \& {Nakar}(2025)}]{Maoz_2024}
{Maoz}, D., \& {Nakar}, E. 2025, \apj, 982, 179, \dodoi{10.3847/1538-4357/ada3bd}

\bibitem[{{McMillan}(2017)}]{McMillan_2017}
{McMillan}, P.~J. 2017, \mnras, 465, 76, \dodoi{10.1093/mnras/stw2759}

\bibitem[{{M{\"u}ller} {et~al.}(2018){M{\"u}ller}, {Gay}, {Heger}, {Tauris}, \& {Sim}}]{Muller_2018}
{M{\"u}ller}, B., {Gay}, D.~W., {Heger}, A., {Tauris}, T.~M., \& {Sim}, S.~A. 2018, \mnras, 479, 3675, \dodoi{10.1093/mnras/sty1683}

\bibitem[{{Noutsos} {et~al.}(2013){Noutsos}, {Schnitzeler}, {Keane}, {Kramer}, \& {Johnston}}]{Noutsos_2013}
{Noutsos}, A., {Schnitzeler}, D.~H.~F.~M., {Keane}, E.~F., {Kramer}, M., \& {Johnston}, S. 2013, \mnras, 430, 2281, \dodoi{10.1093/mnras/stt047}

\bibitem[{{O'Doherty} {et~al.}(2023){O'Doherty}, {Bahramian}, {Miller-Jones}, {Goodwin}, {Mandel}, {Willcox}, {Atri}, \& {Strader}}]{ODoherty_2023}
{O'Doherty}, T.~N., {Bahramian}, A., {Miller-Jones}, J. C.~A., {et~al.} 2023, \mnras, 521, 2504, \dodoi{10.1093/mnras/stad680}

\bibitem[{{Ofek}(2009)}]{Ofek_2009}
{Ofek}, E.~O. 2009, \pasp, 121, 814, \dodoi{10.1086/605389}

\bibitem[{{Podsiadlowski} {et~al.}(2004){Podsiadlowski}, {Langer}, {Poelarends}, {Rappaport}, {Heger}, \& {Pfahl}}]{Podsiadlowski_2004}
{Podsiadlowski}, P., {Langer}, N., {Poelarends}, A.~J.~T., {et~al.} 2004, \apj, 612, 1044, \dodoi{10.1086/421713}

\bibitem[{{Smits} {et~al.}(2006){Smits}, {Maccarone}, {Kundu}, \& {Zepf}}]{Smits:2006}
{Smits}, M., {Maccarone}, T.~J., {Kundu}, A., \& {Zepf}, S.~E. 2006, \aap, 458, 477, \dodoi{10.1051/0004-6361:20065298}

\bibitem[{{Valli} {et~al.}(2025){Valli}, {De Mink}, {Justham}, {Callister}, {Johnston}, {Kresse}, {Langer}, {Rubio}, {Vigna-G{\'o}mez}, \& {Wang}}]{Valli_2025}
{Valli}, R., {De Mink}, S.~E., {Justham}, S., {et~al.} 2025, arXiv e-prints, arXiv:2505.08857, \dodoi{10.48550/arXiv.2505.08857}

\bibitem[{{Van den Heuvel} \& {Van Paradijs}(1997)}]{VandenHeuvel_1997}
{Van den Heuvel}, E.~P.~J., \& {Van Paradijs}, J. 1997, \apj, 483, 399, \dodoi{10.1086/304249}

\bibitem[{{Verbiest} {et~al.}(2012){Verbiest}, {Weisberg}, {Chael}, {Lee}, \& {Lorimer}}]{Verbiest_2012}
{Verbiest}, J.~P.~W., {Weisberg}, J.~M., {Chael}, A.~A., {Lee}, K.~J., \& {Lorimer}, D.~R. 2012, \apj, 755, 39, \dodoi{10.1088/0004-637X/755/1/39}

\bibitem[{{Verbunt} \& {Cator}(2017)}]{Verbunt_Cator_2017}
{Verbunt}, F., \& {Cator}, E. 2017, Journal of Astrophysics and Astronomy, 38, 40, \dodoi{10.1007/s12036-017-9474-5}

\bibitem[{{Verbunt} {et~al.}(2017){Verbunt}, {Igoshev}, \& {Cator}}]{Verbunt_2017}
{Verbunt}, F., {Igoshev}, A., \& {Cator}, E. 2017, \aap, 608, A57, \dodoi{10.1051/0004-6361/201731518}

\bibitem[{{Virtanen} {et~al.}(2020){Virtanen}, {Gommers}, {Oliphant}, {Haberland}, {Reddy}, {Cournapeau}, {Burovski}, {Peterson}, {Weckesser}, {Bright}, {Van der Walt}, {Brett}, {Wilson}, {Millman}, {Mayorov}, {Nelson}, {Jones}, {Kern}, {Larson}, {Carey}, {Polat}, {Feng}, {Moore}, {VanderPlas}, {Laxalde}, {Perktold}, {Cimrman}, {Henriksen}, {Quintero}, {Harris}, {Archibald}, {Ribeiro}, {Pedregosa}, {Van Mulbregt}, \& {SciPy 1. 0 Contributors}}]{Virtanen_2020}
{Virtanen}, P., {Gommers}, R., {Oliphant}, T.~E., {et~al.} 2020, Nature Methods, 17, 261, \dodoi{10.1038/s41592-019-0686-2}

\bibitem[{{Willcox} {et~al.}(2021){Willcox}, {Mandel}, {Thrane}, {Deller}, {Stevenson}, \& {Vigna-G{\'o}mez}}]{Willcox_2021}
{Willcox}, R., {Mandel}, I., {Thrane}, E., {et~al.} 2021, \apjl, 920, L37, \dodoi{10.3847/2041-8213/ac2cc8}

\bibitem[{{Yao} {et~al.}(2017){Yao}, {Manchester}, \& {Wang}}]{Yao_2017}
{Yao}, J.~M., {Manchester}, R.~N., \& {Wang}, N. 2017, \apj, 835, 29, \dodoi{10.3847/1538-4357/835/1/29}

\bibitem[{{Zhao} {et~al.}(2023){Zhao}, {Gandhi}, {Dashwood Brown}, {Knigge}, {Charles}, {Maccarone}, \& {Nuchvanichakul}}]{Zhao_2023}
{Zhao}, Y., {Gandhi}, P., {Dashwood Brown}, C., {et~al.} 2023, \mnras, 525, 1498, \dodoi{10.1093/mnras/stad2226}

\end{thebibliography}
